\documentclass[showpacs,preprintnumbers,amssymb,axodraw,nofootinbib]{revtex4}
\usepackage{amsmath}
\usepackage{graphicx}
\usepackage{epsf}
\usepackage{psfrag}
\usepackage{epsfig}
\usepackage{epstopdf}
\usepackage{graphics}
\usepackage{axodraw4j}
\usepackage[dvips]{color}
\newcommand{\pslash}{\hbox{$p\!\!\!{\slash}$}}

\newcommand{\kslash}{\hbox{$k\!\!\!{\slash}$}}
\newcommand{\lslash}{\hbox{$l\!\!\!{\slash}$}}

\newcommand{\be}{\begin{equation}}
\newcommand{\ee}{\end{equation}}
\newcommand{\bq}{\begin{eqnarray}}
\newcommand{\eq}{\end{eqnarray}}

\begin{document}
	
	\title{Advances towards the systematization of calculations with Implicit Regularization}

	\date{\today}	
	\author{B. Z. Felippe$^{(a,b)}$} \email[] {brunofelippe@unifei.edu.br}	
	\author{A. P. Ba\^eta Scarpelli$^{(a)}$} \email[] {scarpelli@cefetmg.br}
	\author{A. R. Vieira$^{(c)}$} \email[]{alexandre.vieira@uftm.edu.br}
	\author{J. C. C. Felipe$^{(d)}$} \email[]{jean.cfelipe@ufvjm.edu.br}
	
	\affiliation{(a)Centro Federal de Educa\c{c}\~ao Tecnol\'ogica - MG \\
		Avenida Amazonas, 7675 - 30510-000 - Nova Gameleira - Belo Horizonte-MG - Brazil}
	
	\affiliation{(b) Universidade Federal de Itajub\'a - Campus Itabira \\
		Rua Irmã Ivone Drumond, 200 - Distrito Industrial II, Itabira, Minas Gerais, Brazil}
	
	\affiliation{(c) Universidade Federal do Tri\^angulo Mineiro, Campus Iturama, Iturama - MG - Brasil}
	
	\affiliation{(d) Instituto de Engenharia, Ci\^encia e Tecnologia, Universidade Federal dos Vales do Jequitinhonha e Mucuri, Avenida Um, 4050 - 39447-790 - Cidade Universit\'aria - Jana\'uba - MG- Brazil}
	
	\begin{abstract}
		\noindent
		There is currently a high demand for theoretical predictions for processes at next-to-next-to-leading order (NNLO) and beyond, mainly due to the large amount of data which has already been collected at LHC. This requires practical methods that meet the physical requirements of the models under study. We develop a new procedure for applying Constrained Implicit Regularization which simplifies the calculation of amplitudes, including finite parts. The algebraic identities to separate the divergent parts free from the external momenta are used after the Feynman parametrization. These algebraic identities establish a set of scale relations which are always the same and do not need to be calculated in each situation. This procedure unifies the calculations in massive and non-massive models in an unique procedure. We establish a systematization of the calculation of one-loop amplitudes and extend the procedure for higher-loop orders.
		
	\end{abstract}
	
	\pacs{11.10.Gh, 11.30.-j, 11.15.-q}

	\maketitle
	
	\section{Introduction}
	\label{Introduction}
	The success of Quantum Field Theory is highly attached to the principles of renormalization, which allow, through suitable renormalization conditions, the redefinition of the quantities of interest in order to obtain physical meaningful results. This is because the perturbative calculations of transition amplitudes involve divergences. In the high energy regime, one has to deal with ultraviolet (UV) divergences coming from the locality of fundamental interactions. The calculation of Feynman amplitudes, thus, is intricate with the technical problem of regularizing integrals in a consistent way such that the finite part presents the expected physical content of the amplitude.
	
	The regularization technique to be used should meet some important features of the $S$-matrix, like the preservation of unitarity and causality, the preservation of gauge symmetry and supersymmetry, among others. Besides, the renormalization procedure can be more involved depending on the regularization applied. The usual textbook technique is Dimensional Regularization \cite{DimReg}, which is powerful due to gauge symmetry preservation in all orders of perturbation theory. The dimensional extension, however, causes difficulties when the model under investigation encompasses dimensional dependent mathematical objects, like the $\gamma_5$ matrix. This is the case of topological and supersymmetric theories. Dimensional Reduction \cite {DimRed} is a way out of this problem for extending only the dimension of the Feynman integrals, but preserving the symmetry group algebra in the original space-time dimension. However, there are some mathematical inconsistencies in the procedure when the calculation is performed beyond one-loop order. The quantum action principle has been used in order to consistently apply Dimensional Reduction beyond one-loop order in supersymmetric models \cite{Stockinger}. These difficulties with dimensional extension are the motivating fact for several developments of regularization procedures which are carried out in the proper dimension of the model (good discussions on this topic are carried out in \cite{Marcos-1}, \cite{Marcos-2} and \cite{NNLO}).
	
	Differential Renormalization (DR) is one of these approaches \cite{DR}. It works in the proper dimension of the theory in coordinate space, and has been proved to be simple and powerful in many applications \cite{DR2}-\cite{DR5}. It consists in the manipulation of singular distributions attributing to them properties of the regular ones. In the end, these singularities are substituted by renormalized functions and several mass parameters are introduced in the results. Relations between these parameters are established in order to preserve symmetries. A further development in order to automatically satisfy symmetries came with the constrained version of Differential Renormalization (CDR) \cite{cdf1}-\cite{cdf8}.
	
	Implicit Regularization \cite{ir1}-\cite{ir3}, on the other hand, is carried out in momentum-space, also in the physical dimension of the theory. The basic idea of the Implicit Regularization (IReg) procedure of a Feynman integral is to consider, before manipulating the integrands, the presence of some implicit regularization scheme or function. The scheme composes the originally divergent integral and allows the separation of its part dependent on the regularization from the finite part, which must be independent of the regularization used. The separation can be done by applying a simple algebraic identity to the integrand, such that the divergent parts are written only in terms of the internal momentum in the loops and do not need to be evaluated. The independence of the divergent integrals from the external momentum is a highly desirable feature, since we will only need local counterterms in the Lagrangian of the model in order to eliminate any divergences that arise in the perturbative calculus. Furthermore, these divergent integrals can be written as functions of an arbitrary mass parameter that characterizes the freedom of separating of the divergent part of an amplitude and plays the role of scale in the renormalization group equation. Symmetries of the model or phenomenological requirements determine arbitrary parameters which are surface terms and arise from the procedure. There is a special choice for the parameters that automatically delivers symmetric amplitudes in anomaly free cases. Fixing these parameters at the beginning of the calculation considerably simplifies the application of the method. This results in a constrained version of Implicit Regularization (CIReg). Implicit Regularization has been successfully applied to a wide variety of problems, including non-abelian and supersymmetric models, and calculations beyond the one-loop order \cite{ir4}-\cite{irn}. There is a class of four dimensional regularization techniques, however, CIReg included, for which the difficulties with mathematical objects like the $\gamma_5$ matrix and the Levi-Civita tensor remain. One deals, for example, with ambiguities in the Dirac trace involving the $\gamma_5$ matrix. This problem was discussed in \cite{Bruque}, in which a consistency procedure was proposed.  
	
	It is important to note that, in its traditional formulation, the procedure of IReg is applied after performing the algebra of the symmetry group in the amplitude, which is decomposed into a combination of integrals. Each integral, then, is separated into finite and divergent parts. The Feynman parametrization process is applied only to finite integrals. However, for a single Feynman integral, the expansion to separate the divergences generates a set of finite integrals, which can be large in case of high degrees of divergence. In addition, expansions generate high powers of momenta in the numerator and denominator, which can complicate the calculations quite a bit.
	
	The technique known as Loop Regularization (LORE) \cite{Lore-1}-\cite{Lore-3} has some aspects in common with Constrained Implicit Regularization, such as the use of consistency conditions, which, in practice, eliminate surface terms that cause violation of symmetries. LORE prescribes that Feynman parametrization is applied to the amplitude as a whole, like in Dimensional Regularization. Afterwards, the algebra of the symmetry group is performed and then the integration in momenta. In the case of divergent parts, consistency conditions are used to write the amplitude in terms of scalar loop integrals. Finally, these scalar integrals are regularized similarly to the Pauli-Villars procedure and then calculated.
	
	In this paper, we propose a new approach for implementing  Implicit Regularization which greatly simplifies the calculation of the amplitudes, mainly the finite parts. Assuming the action of a regularization, we apply, as in LORE, Feynman parametrization in the complete amplitude and eliminate the surface terms through what we call consistency relations, so as to have only scalar divergent integrals. We then make use of algebraic identities to expand the integrands to obtain basic divergences which are free from the external momenta. These algebraic identities establish a set of scale relations which are always the same and do not need to be calculated in each situation. As a byproduct, the scale relations allow to introduce an arbitrary mass scale that will be useful in the process of renormalization. The results of this new approach are the same of Constrained Implicit Regularization. However, the procedure also permits the introduction of local arbitrary parameters for the models they are needed. This approach is also extended for multiloop calculations, since there is currently a high demand for theoretical predictions for processes at next-to-next-to-leading order (NNLO) and beyond, mainly due to the large amount of data which has already been collected at LHC. Therefore, practical methods for higher order calculations are being intensively investigated.
	
	The paper is divided as follows: in section II we present a brief review of the traditional procedure for Implicit Regularization, with examples to be compared with the new ones; in section III, the new for implementing constrained IReg is presented, with discussions and comparisons with the cases of the previous section; in section IV, we carry out a systematization of the calculation of one-loop amplitudes; in section V, we present the extension for higher-loop calculations; discussions on implicit regularization of infrared divergences are carried out in section VI; section VII is left for conclusions and perspectives.
	
	\section{A brief review of Implicit Regularization}
	
	Implicit Regularization (IReg) can be formulated by a set of rules. The first thing to be done is to assume a regularization is applied to the complete amplitude, so as algebraic manipulations can be carried out in the integrand. We then perform the group algebra and write the momentum-space amplitude as a combination of basic integrals, multiplied by polynomials of the external momentum and typical objects of the symmetry group. We give below examples of basic integrals:
	\be
	\label{Integrals}
	I,I_\mu, I_{\mu\nu}= \int^R \frac{d^4k}{(2\pi)^4}\frac{1,k_\mu,k_\mu k_\nu}{(k^2-m^2)[(p-k)^2-m^2]},
	\ee
	in which the index $R$ in the integrals is to indicate they are regularized.  Each one of these basic integrals can be treated following a set of rules. So, a table with their results can be used whenever a new
	calculation is being performed. The rules of Constrained Implicit Regularization (CIReg) for calculations at one-loop order can be stated as:
	
	\begin{enumerate}
		
		\item an amplitude is assumed to be regularized with a 4D technique which is maintained implicit;
		
		\item to obtain the divergent part of a basic integral, we apply recursively the identity,
		\bq
		\frac {1}{(p-k)^2-m^2}=\frac{1}{(k^2-m^2)}
		-\frac{p^2-2p \cdot
			k}{(k^2-m^2)
			\left[(p-k)^2-m^2\right]},
		\label{ident}
		\eq
		so as the divergent part do not have the external momentum $p$ in the denominator. This will assure local counterterms. The remaining divergent integrals have the form
		\be
		\int_k ^R \frac{k_{\mu_1}k_{\mu_2}\cdots}{(k^2-m^2)^\alpha},
		\ee
		in which we use $\int_k$ as a simplification of $\int d^4k/(2 \pi)^4$;
		
		\item  The divergent integrals with Lorentz indices must be expressed in terms of divergent scalar integrals and surface terms. For example:
		\be
		\int_k ^R \frac{k_\mu k_\nu}{(k^2-m^2)^3}=
		\frac 14 \left\{ \eta_{\mu \nu} \int_k^R
		\frac {1}{(k^2-m^2)^2} -\int_k^R \frac{\partial}{\partial k^\nu}
		\left( \frac{k_\mu}{(k^2-m^2)^2}\right)\right\}.
		\label{ts}
		\ee
		
		The surface terms, that vanish for finite integrals, depend here on the method of regularization to be applied. They are symmetry-violating terms, since the possibility of making shifts in the integrals needs the surface terms to vanish. Non-null surface terms imply that the amplitude depend on the momentum routing choice. In practice, setting them zero from the beginning is equivalent to canceling these surface terms by means of local symmetry-restoring counterterms;

		\item Finally, the divergent part of the integrals is written as a combination of the basic divergences
		\be
		I_{log}(m^2)=\int_k^R \frac{1}{(k^2-m^2)^2}
		\,\,\,\,\,\,\,\,  \mbox{and}  \,\,\,\,\,\,\,\
		I_{quad}(m^2)=\int_k^R \frac{1}{(k^2-m^2)},
		\label{basicdiv}
		\ee
		which will require local counterterms in the process of renormalization.
		
	\end{enumerate}
	
	Let us comment on the first item above. Ideally, the regularization technique that is maintained implicit should have the properties of not modifying neither the integrand nor the dimension of spacetime. The former property is to preserve the finite part and the latter is a requirement in order to not violate supersymmetry. In previous works, the cutoff regularization had been cited as a possibility, although it is known that it causes the violation even of simple symmetries. This should be understood in the following sense. When surface terms are fixed to a given value (zero, for example) in order to obtain a symmetric result, this is completely equivalent to add symmetry restoring counterterms to the amplitude. If we expand the Feynman integral as prescribed by IReg and then solve the divergent parts by using cutoff regularization, the symmetry violating terms will come from the tensorial divergent basic integrals. Thus, the procedure of IReg allows us to identify all the symmetry-violating terms. We know exactly which part should be subtracted. In this sense, as long as the implicit technique is defined in the proper dimension of the model and do not modify the integrand, it will not get in the way of the procedure. In practice, we do not make use of any specific regularization. It was shown in papers as \cite{bphz1} and \cite{bphz2}, the connection of the method with the Bogoliubov–Parasiuk–Hepp–Zimmermann (BPHZ) theorem. Therefore, the IReg can also be seen as a map of Feynman amplitudes into results with desirable properties.

	Let us now present an example to show how the traditional procedure of IReg applies to a complete amplitude in order to compare, in the next section, with the new proposed approach. We consider below the vacuum polarization tensor of spinorial QED, which after the use of Feynman rules, is given by
	\be
	-i \Pi^{\mu\nu}=q^2 \int_k^R \frac{\mbox{tr}\left\{\gamma^\mu (\kslash - \pslash + m)\gamma^\nu(\kslash + m)\right\}}{(k^2-m^2)[(k-p)^2-m^2]}.
	\label{vpt}
	\ee
	
	Following the steps listed above, we first calculate the trace and write the amplitude as a combination of basic integrals. We end up with
	\be
	-i \Pi^{\mu\nu}= 4 q^2 \left\{2 I^{\mu\nu} - p^\mu I^\nu - p^\nu I^\mu - \frac 12 \eta^{\mu\nu}\left[I_{quad}(m^2) + I_1 - p^2 I\right] \right\},
	\ee
	where $I$, $I_\mu$ and $I_{\mu\nu}$ are defined in eq. (\ref{Integrals}), $I_{quad}(m^2)$ is given in (\ref{basicdiv}) and
	\be
	I_1 = \int_k^R \frac{1}{[(k-p)^2-m^2]}.
	\ee
	
	The next step is to calculate each one of the integrals. Let us give the directions of the calculations of $I_{\mu\nu}$. It is a divergent integral and the integrand need to be expanded using (\ref{ident}) in order to separate the finite part from the regularization dependent one. We have, after discarding the null integrals,
	\bq
	&&I_{\mu\nu}= \int_k^R \frac{k_\mu k_\nu}{(k^2-m^2)^2} - p^2 \int_k^R \frac{k_\mu k_\nu}{(k^2-m^2)^3} 
	+ 4 p^\alpha p^\beta \int_k^R \frac{k_\mu k_\nu k_\alpha k_\beta}{(k^2-m^2)^4} + \nonumber \\
	&& +p^4 \int_k \frac{k_\mu k_\nu}{(k^2-m^2)^4} - \int_k \frac{[p^2-2(p \cdot k)]^3 k_\mu k_\nu}{(k^2-m^2)^4[(k-p)^2-m^2]}.
	\label{imunu-expand}
	\eq
	The first term is quadratically divergent and the second and third terms are logarithmically divergent. We use the procedure of (\ref{ts}) to obtain scalar basic divergences and follow the notation from reference \cite{A-Vieira}:
	\be
	\int_k^R \frac{k_\mu k_\nu}{(k^2-m^2)^2}= \frac{\eta_{\mu\nu}}{2}\left[I_{quad}(m^2) - \upsilon_2 \right],
	\label{TS-quad}
	\ee
	\be
	\int_k^R \frac{k_\mu k_\nu}{(k^2-m^2)^3}= \frac{\eta_{\mu\nu}}{4}\left[I_{log}(m^2) - \upsilon_0 \right]
	\label{TS-0}
	\ee
	and
	\be
	\int_k^R \frac{k_\mu k_\nu k_\alpha k_\beta}{(k^2-m^2)^4}= \frac{\eta_{\mu\nu\alpha\beta}}{24}\left[I_{log}(m^2) - \xi_0 \right],
	\ee
	in which $\eta_{\mu\nu\alpha\beta}\equiv\eta_{\mu\nu}\eta_{\alpha\beta}+\eta_{\mu\alpha}\eta_{\nu\beta}+\eta_{\mu\beta}\eta_{\nu\alpha}$ and 
	where $\upsilon_0$, $\upsilon_2$ and $\xi_0$ are surface terms. These terms are arbitrary and regularization dependent because they are 
	differences between two integrals with the same degree of divergences, as shown in eq. (\ref{ts}). In this case, the indices $0$ and $2$ corresponds to logarithmic and quadratic divergences, respectively.
	
	The last two integrals in (\ref{imunu-expand}) are finite and can be solved. The Feynman parametrization can be applied when necessary, like in the last one. Note that the high power on the momenta in the numerator and in the denominator make calculations longer. The final result for $I_{\mu\nu}$ is given by
	\bq
	I_{\mu\nu}&=&\frac{\eta_{\mu\nu}}{2}I_{quad}(m^2) 
	+ \frac {1}{12}\left(-p^2 \eta_{\mu\nu} + 4 p_\mu p_\nu\right) I_{log}(m^2)
	- \frac{\eta_{\mu\nu}}{12}(6\upsilon_2 -3p^2\upsilon_0 +2p^2 \xi_0 )- \frac{p_\mu p_\nu}{3} \xi_0 + \nonumber \\
	&+& \frac{i}{(4\pi)^2} \left\{\frac{1}{12p^2}\left[ (p^2-4m^2)p^2 \eta_{\mu\nu}
	-4(p^2-m^2) p_\mu p_\nu \right] Z_0(p^2,m^2,m^2,m^2) + \right. \nonumber \\
	&+& \left. \frac{1}{18}(p_\mu p_\nu - p^2 \eta_{\mu\nu})\right\},
	\eq
	where
	\be
	Z_k(p^2,m_1^2,m_2^2,m_3^2)=\int_0^1 dz\,\, z^k \ln{\left\{\frac{p^2z(1-z)+(m_1^2-m_2^2)z-m_1^2}{(-m_3^2)}\right\}}.
	\ee
	The same procedure is used for calculating the other Feynman integrals. We obtain, for the vacuum polarization tensor,
	\bq
	-i \Pi^{\mu\nu}&=&\frac 43 (p^2\eta^{\mu\nu} - p^\mu p^\nu)\left\{I_{log}(m^2)
	-\frac{i}{(4\pi)^2}\left[\frac{(p^2+2m^2)}{p^2}Z_0(p^2,m^2)+\frac 13\right]\right\} + \nonumber \\
	&-&4 \upsilon_2 \eta^{\mu\nu} + \frac 43 \left\{\upsilon_0 (p^2\eta^{\mu\nu} - p^\mu p^\nu)
	- (2p^\mu p^\nu + p^2 \eta^{\mu\nu})(\xi_0 -2\upsilon_0 )\right\},
	\eq
	in which, for economy, we used $Z_k(p^2,m^2,m^2,m^2)\equiv Z_k(p^2,m^2)$. Note that the amplitude is transversal if $\upsilon_2=0$ and $\xi_0=2\upsilon_0$. This approach of Implicit Regularization, in which the surface terms are parametrized and fixed in the end is useful when the model under investigation presents ambiguities as in the cases of chiral anomalies or topological field theories \cite{Viglioni}. The Constrained Implicit Regularization, on the other hand, fixes the surface terms to zero from the beginning. This automatically delivers symmetric results, as it has been shown for Abelian and non-Abelian gauge theories and for supersymmetric models.
	
	\section{A new approach for implementing Constrained Implicit Regularization}
	\label{new-approach}
	
	The basic idea of Constrained Implicit Regularization, as we already discussed, is to assume the presence of a regularization with the aim of using mathematical identities in order to separate the regularization dependent from the finite part. The divergent part is a combination of scalar basic divergences, which are obtained after the use of consistency relations that eliminate surface terms. Here, we propose a procedure for applying Constrained Implicit Regularization which simplifies enormously the process of calculation. The steps are listed below:
	\begin{enumerate}
		\item as in the original procedure, a regularization scheme is assumed to be acting in the complete amplitude, with the same desirable characteristics already presented;
		\item Feynman parametrization is applied to the Feynman integrals. It can be applied to the complete amplitude. When it is carried out in the complete amplitude, the needed shift in the momentum of integration is just a modification in the loop momentum;
		\item the algebra of the group of symmetry is carried out. In the case the integrals are treated separately, the algebra is performed before Feynman parametrization;
		\item the integrals in the momenta are separated by degree of divergence, all with even powers of the integration momentum in the numerator, of the type
		\be
		\int^R\frac{d^4k}{(2\pi)^4}\frac{1, k_\mu k_\nu, k_\mu k_{\nu} k_{\alpha}k_{\beta},\cdots}{(k^2+H^2)^n},
		\ee
		being $H^2$ function of the external momenta, of the masses and of the Feynman parameters. If factors of $k^2$ appear in the numerator, they should be canceled with factors in the denominator by adding and subtracting $H^2$;
		\item for the divergent parts, the surface terms are eliminated by means of the consistency relations, in order to obtain scalar integrals. For one-loop logarithmically and quadratically divergent integrals, we set, respectively,
		\be
		\int^R\frac{d^4k}{(2\pi)^4}\frac{k_{\mu_1} k_{\mu_2} \cdots k_{\mu_n}}{(k^2+H^2)^{2+\frac n2}}=
		\frac{\eta_{\mu_1 \mu_2 \cdots \mu_n}}{2^{\frac n2} (\frac n2 +1)!}\left\{ 
		\int^R\frac{d^4k}{(2\pi)^4}\frac{1}{(k^2+H^2)^2} - \alpha_{\frac n2}   \right\}
		\label{RC-geral}
		\ee
		and
		\be
		\int^R\frac{d^4k}{(2\pi)^4}\frac{k_{\mu_1} k_{\mu_2} \cdots k_{\mu_n}}{(k^2+H^2)^{1+\frac n2}}=
		\frac{\eta_{\mu_1 \mu_2 \cdots \mu_n}}{2^{\frac n2} (\frac n2)!}\left\{ 
		\int^R\frac{d^4k}{(2\pi)^4}\frac{1}{(k^2+H^2)} - \beta_{\frac n2}   \right\},
		\label{RC-geral-2}
		\ee
		in which $n$ is even and $\eta_{\mu_1 \mu_2 \cdots \mu_n}$ is the symmetric combination of the products of metric tensors, $\eta_{\mu_1 \mu_2}\cdots \eta_{\mu_{n-1}\mu_n}$, with coefficient $1$. In the expression above, we left the parameters for the surface terms $\alpha_{\frac n2}$ and $\beta_{\frac n2}$ only for completeness, since in the constrained version of IReg they are fixed null. In order to obtain the relations above, we use recursively the relation,
		\be
		\int_k^R \frac{\partial}{\partial k^{\mu_n}}\left(\frac{k_{\mu_1}\cdots k_{\mu_{n-1}}}{(k^2+H^2)^{m-1}}\right) =
		\int_k^R \frac{{\cal S}[\eta_{\mu_1 \mu_n}k_{\mu_2}\cdots k_{\mu_{n-1}}]}{(k^2+H^2)^{m-1}}
		- 2 (m-1) \int_k^R\frac{k_{\mu_1} \cdots k_{\mu_n}}{(k^2+H^2)^m},
		\ee
		until the first integral of the second member of the equation is scalar. Apart two integrals, including the 
		scalar one, all the others will be surface terms and are gathered in one parameter which will be fixed null. In the equation above, we define 
		${\cal S}[T_{\mu_1 \cdots \mu_n}]$ as the minimal symmetrization of the tensor $T$, in the sense that only distinct 
		terms are considered and all of them have coefficient one. For example, ${\cal S}[k_\mu k_\nu p_\alpha]= k_\mu k_\nu 
		p_\alpha + k_\mu k_\alpha p_\nu + k_\nu k_\alpha p_\mu$. Important to note that $m=\frac n2 + 2$ for logarithmic 
		divergences and $m=\frac n2 + 1$ for quadratic divergences. By the definitions of equation (\ref{basicdiv}), the 
		remaining scalar divergences above are $I_{log}(-H^2)$ and $I_{quad}(-H^2)$.
		\item in complement to the previous step, since in CIR surface terms are set to zero, shifts in divergent integrals are allowed;
		\item the next step, which is one of the basic ideas of IReg, is the use of algebraic identities in order to 
		get the divergent integrals free from the external momenta. Here, we use recursively a simpler expansion,
		\be
		\frac{1}{(k^2+H^2)}= \frac{1}{(k^2-\lambda^2)} - \frac{\lambda^2+H^2}{(k^2-\lambda^2)(k^2+H^2)}.
		\ee
		We have the advantage of obtaining closed expressions to be used in any calculation:
		\be
		I_{log}(-H^2)= I_{log}(\lambda^2) - \frac{i}{16\pi^2}\ln{\left(-\frac{H^2}{\lambda^2}\right)}
		\label{scale-1}
		\ee
		and
		\be
		I_{quad}(-H^2)=I_{quad}(\lambda^2)-(\lambda^2+H^2)I_{log}(\lambda^2) - \frac{i}{16\pi^2} 
		\left[\lambda^2 + H^2 - H^2 \ln{\left(-\frac{H^2}{\lambda^2}\right)} \right].
		\label{scale-2}
		\ee
		These are called scale relations, which, as a byproduct, introduce an energy scale for the renormalization group, $\lambda^2$, that can be simply one of the masses of the model. The basic divergences are now factorized out of the integrals in the Feynman parameters, which can be computed. As in the traditional formulation of IReg, the divergent part of the amplitudes is written in terms of these basic divergences: $I_{log}(\lambda^2)$, $I_{quad}(\lambda^2)$, etc.
	\end{enumerate} 
	
	The implementation of the above steps simplifies a lot the calculations of the finite parts. An additional advantage is related to models which present fields with different masses or non-massive fields. In non-massive models, as long as the off-shell amplitude is infrared finite, the traditional procedure requires that a fictitious mass is introduced in the propagator so as the expansion in the integrand can be carried out. At the end of the calculation, the scale relations are used to remove the fictitious mass from the divergent part so that the limit of the mass going to zero can be taken. Here, the procedure is unified, since all the mass dependence is inside $H^2$.
	
	It is important to carry out a discussion on the possibility of applying Feynman parametrization before the use of the identities of Implicit Regularization. Normally, in regularization processes, the amplitude is regularized and, then, considering that the integral is well defined, the Feynman parametrization is carried out. In Dimensional Regularization, for example, the amplitude is extended to dimension $d$ and, after this procedure, the parametrization is promptly applied. The key point in Feynman parametrization is the shift, $k \to k + f(p_i,x_i)$ ($f(p_i,x_i)$ is a linear function of the external momenta $p_i$ and of the parameters $x_i$),  which is carried out to solve the $k$ integral. When the integral is assumed to be finite, this shift is allowed, since it will not generate any surface term. In the case of IReg, there is not an explicit regularization applied. So, in the more general approach of IReg, care is taken that surface terms are not discarded. Therefore, expansions are performed on the integrand in order to separate the divergent parts from the finite ones. The Feynman parametrization is then applied only to the finite part, which will not produce surface terms. The surface terms are thus identified and parametrized in the consistency relations used to reduce the tensorial divergent basic integrals to scalar ones. Part of the surface terms that are maintained and parametrized in the more general IReg is related with the momentum routing in the loops. When Feynman parametrization is performed in the whole amplitude before the expansion of IReg is carried out, this part is lost, as Feynman's post-parametrization shift just eliminates the routing dependency, as it is evident in the following calculation. Therefore, the steps described above exactly implement the Constrained Implicit Regularization, in which all the surface terms are set to zero. The identities used in constrained IReg, then, commute with the Feynman parametrization of divergent integrals. 
	
	Just as an example, we show below, for a simple linearly divergent integral, the equivalence of applying the Feynman parameterization before or after the expansion of the integrals. Let us consider the massless integral $I_\mu$ and its known result within Implicit Regularization,
	\be
	I_\mu = \int_k^R \frac{k_\mu}{k^2(p-k)^2} = \frac{p_\mu}{2} \left\{I_{log}(\lambda^2) + 
	\upsilon_0 + \frac{i}{16 \pi^2}\left[2 - \ln\left(-\frac{p^2}{\lambda^2}\right)\right]\right\},
	\ee
	being $\upsilon_0$ the surface term defined in (\ref{TS-0}). We now solve this same integral by applying Feynman parametrization before. We have
	\be
	I_\mu = \int_0^1 \, dx \, \int_k^R \frac{k_\mu}{\left[(k-px)^2 + H^2\right]^2},
	\label{k-integral}
	\ee
	with $H^2=p^2x(1-x)$, and in which we avoided to carry out the traditional shift. Now the integral in momentum should be compared with the also known result within Implicit Regularization,
	\bq
	&&I_\mu(p_1,p_2,m^2)=\int_k^R \frac{k_\mu}{[(k-p_1)^2-m^2][(k-p_2)^2-m^2]}
	=\frac{(p_1 + p_2)_\mu}{2} \left\{ I_{log}(m^2) + \upsilon_0 + \right.\nonumber \\
	&& - \left. \frac{i}{16 \pi^2} 
	\int_0^1 \, du \, \ln \left[ \frac{(p_1-p_2)^2 u(1-u) - m^2}{(-m^2)}\right]\right\}.
	\eq
	We then identify the $k$ integral of (\ref{k-integral}) with
	\bq
	I_\mu(xp,xp,-H^2) &=& x p_\mu \left\{ I_{log}(-H^2) + \upsilon_0 
	- \frac{i}{16 \pi^2} \int_0^1 \, du \, \ln \left( \frac{ H^2}{H^2}\right)\right\} 
	\nonumber\\
	&=& x p_\mu \left[ I_{log}(-H^2) + \upsilon_0 \right].
	\eq
	Substituting this in (\ref{k-integral}), we have
	\bq
	I_\mu &=& \int_0^1 \, dx \,x p_\mu\left[ I_{log}(-H^2) + \upsilon_0 \right] 
	= \int_0^1 \, dx \,xp_\mu\left[ I_{log}(\lambda^2) - \frac{i}{16 \pi^2} \ln\left(-\frac{H^2}{\lambda^2} \right) + \upsilon_0 \right] \nonumber \\
	&=& \frac{p_\mu}{2} \left\{I_{log}(\lambda^2) + 
	\upsilon_0 + \frac{i}{16 \pi^2}\left[2 - \ln\left(-\frac{p^2}{\lambda^2}\right)\right]\right\},
	\eq
	that is the same result as in the calculation in which the Feynman parametrization is carried out only after the expansion. It is important to note that the effect of making the shift $k \to k + px$ in (\ref{k-integral}) is the automatic discard of the surface term $\upsilon_0$. 
	
\begin{figure}[h]
		\begin{center}
			\includegraphics[scale=1]{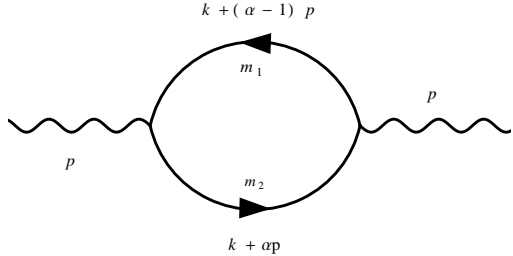}
			\caption{Diagram which contributes to the self-energy of a massive vector-field. The fermionic fields in the loop have different masses and the momentum routing in the amplitude reads $k+(\alpha - 1)p$ for the propagator with mass $m_{1}$ and $k+\alpha p$ for the propagator with mass $m_{2}$. The wavy and solid lines represent the vectorial and fermionic fields, respectively.}
			\label{Fig1}
		\end{center}
	\end{figure}

	Let us now carry out an example of calculation which is a little more involved than the QED vacuum polarization tensor and consider the vector-field self-energy in which the two fermions in the loop have different masses. It is the case of the influence of the heavy quarks, the doublet $(t,b)$, in the corrections to the $W$-boson mass (see, for example, \cite{Hollik}). The corresponding Feynman graph is depicted in Figure \ref{Fig1}. Let us also, for pedagogical reasons, assign an arbitrary distribution of momenta in the internal lines. The amplitude is proportional to
	
	\be
	\Pi^{\mu\nu}(p,m_1,m_2)=\int_k^R \frac{\mbox{tr}\left\{\gamma^\mu [\kslash + (\alpha-1) \pslash + m_2]\gamma^\nu[\kslash + \alpha \pslash +  m_1]\right\}} {[(k+\alpha p)^2-m_1^2]\{[k+(\alpha-1)p]^2-m_2^2\}},
	\label{vpt}
	\ee
	which after Feynman parametrization, with the shift $k \to k + (x-\alpha)p$, reads
	\be
	\Pi^{\mu\nu}(p,m_1,m_2)= \int_0^1 dx \, \int_k^R \frac{\mbox{tr}
		\left\{\gamma^\mu[\kslash+(x-1)\pslash + m_2]\gamma^\nu[\kslash +x\pslash + m_1]\right\}}{(k^2+H^2)^2},
	\ee
	with $H^2=p^2x(1-x)+(m_1^2-m_2^2)x-m_1^2$. We note that the entire dependence on the $\alpha$ parameter disappeared. In other words, when the total amplitude is Feynman parametrized together, the amplitude is automatically momentum-routing invariant. After calculating the trace, with only even terms in $k^\mu$ remaining, and canceling terms in $k^2$ by adding and subtracting $H^2$, we stay with
	\bq
	&&\Pi^{\mu\nu}(p,m_1,m_2) = 4 \int_0^1 dx \, \left\{-\eta^{\mu\nu}\int_k^R \frac{1}{k^2+H^2} + 2 \int_k^R \frac{k^\mu k^\nu}{(k^2+H^2)^2} + \right. \nonumber \\
	&& + \left. \left\{ 2(p^2 \eta^{\mu\nu}-p^\mu p^\nu)x(1-x) +(m_1 - m_2) [(m_1+m_2)x-m_1]\eta^{\mu\nu}\right\}\int_k^R \frac{1}{(k^2+H^2)^2}\right\}.
	\eq
	The first two integrals, which are quadratically divergent, cancel out if we use equation (\ref{TS-quad}) and fix $\upsilon_2=0$, as it is prescribed by constrained IReg. For the remaining $I_{log}(-H^2)$, we use the scale relation (\ref{scale-1}), so that the amplitude is written as
	\bq
	\Pi^{\mu\nu}(p,m_1,m_2) &=& 4 \int_0^1 dx \, \left\{ 2(p^2 \eta^{\mu\nu}-p^\mu p^\nu)x(1-x) +(m_1 - m_2) [(m_1+m_2)x-m_1]\eta^{\mu\nu}\right\} \nonumber \\
	&\times&\left\{I_{log}(\lambda^2) - \frac{i}{16 \pi^2} \ln{\left(-\frac{H^2}{\lambda^2}\right)}\right\}.
	\eq
	Finally, we obtain
	\bq
	&&\Pi^{\mu\nu}(p,m_1,m_2)= 8 (p^2\eta^{\mu\nu} - p^\mu p^\nu)\left\{\frac 16 I_{log}(\lambda^2)
	-\frac{i}{(4\pi)^2}(\tilde{Z}_1-\tilde{Z}_2)\right\} \nonumber \\
	&& + 4 (m_1 - m_2) \eta^{\mu\nu}\left\{ -\frac 12 (m_1 - m_2) I_{log}(\lambda^2)
	-\frac{i}{(4\pi)^2}[(m_1 + m_2)\tilde{Z}_1- m_1 \tilde{Z}_0]\right\},
	\eq
	where $\tilde{Z}_k$ is a short for $Z_k(p^2,m_1^2,m_2^2,\lambda^2)$. Since the vector field is massive, the polarization tensor is not transverse. Rather, the amplitude obeys the relation
	\bq
	p_\nu \Pi^{\mu\nu}&=& 4 (m_1 - m_2) p^\mu\left\{ -\frac 12 (m_1 - m_2) I_{log}(\lambda^2)
	-\frac{i}{(4\pi)^2}[(m_1 + m_2)\tilde{Z}_1- m_1 \tilde{Z}_0]\right\} \nonumber \\
	&=& (m_1 - m_2) T^\mu,
	\label{eqWI}
	\eq
	where
	\be
	T^\mu=\int_k^R \frac{\mbox{tr}\left\{\gamma^\mu (\kslash - \pslash + m_2)(\kslash + m_1)\right\}}{(k^2-m_1^2)[(k-p)^2-m_2^2]}.
	\ee	
	This is a very direct and compact calculation. Note that the procedure would be identical in the case of a off-shell calculation in non-massive QED, with the obvious modification of $H^2$, with only the transverse part remaining. 
	
\begin{figure}[h]
		\begin{center}
			\includegraphics[scale=1]{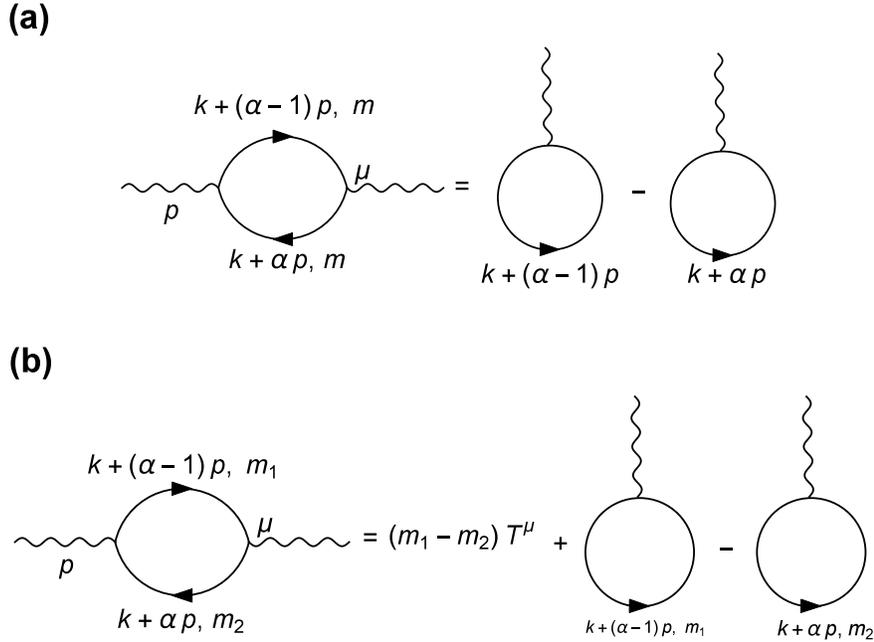}
			\caption{(a) Diagrammatic representation of the gauge and momentum routing invariance relation for QED. (b) The same diagrammatic representation for fermions with different masses.}
			\label{figMRIGI}
		\end{center}
	\end{figure}

	The Ward identity in eq. (\ref{eqWI}) is a particular case of a diagrammatic relation, as depicted in Figure \ref{figMRIGI}. This diagrammatic relation is 
	respected as long as the regularization applied is compatible with shifts in the momentum of integration (see a good discussion in \cite{Bruque}). In the abelian case, QED for example, gauge invariance is fulfilled if and only if momentum routing invariance is as well, as we can 	easily see in Figure \ref{figMRIGI}$(a)$. In the approach we present in this work, the shifts we perform in Feynman parametrization is already an assumption of momentum routing invariance. This already delivers gauge invariant results. Furthermore, when considering the electroweak theory as 	a whole, we have an additional relation as in Fig. \ref{figMRIGI}$(b)$ due to the change of flavors and we recover QED when $m_1=m_2=m$. In this case, mass terms have already broken the larger gauge symmetry $SU(2)\otimes U(1)$ into simply $U(1)$.
	
	It is important to comment on the fact that some aspects of our approach are similar to procedures adopted in Loop Regularization (LORE). In the case of LORE, after the amplitude is Feynman parametrized, consistency conditions are applied which, in practice, discard surface terms. Such conditions, which are the same of IReg, were determined, in that case, by the requirement that symmetries be respected in specific amplitudes \cite{Lore-1} and then generalized. The remaining scalar divergent loop integrals, however, are calculated using a procedure which is similar to the Pauli-Villars regularization \cite{Pauli-Villars}. On the other hand, Implicit Regularization is based on the elimination of surface terms and in the expansion of the integrand so as the renormalization needs only local counterterms. The remaining divergent integrals do not need be explicitly calculated. 
	
	In the next section, we present a systematization for calculations of general one-loop divergent amplitudes.
	
	\section{A systematization for the calculation of one-loop integrals}
	
	We present now a systematization of the calculation of one-loop Feynman integrals in the framework of this new approach for implementing Implicit Regularization. The procedure is very interesting, since the results encompass the finite parts. The methodology we carry out in this section applies to integrals rather than in the complete amplitude. It is not a problem, since different Feynman parametrization in the integrals that are part of the amplitude has implications only in the surface terms which are fixed zero in constrained IReg. 
	
	Let us begin with a general one-loop integral with logarithmic degree of divergence, which is written as
	\be
	I^{(0)}_{\mu_1 \cdots \mu_n}=\int_k^R \frac{k_{\mu_1}\cdots k_{\mu_n}}{(k^2-m^2)[(p_1-k)^2-m_1^2]
		\cdots [(p_r-k)^2-m_r^2]},
	\label{I-zero}
	\ee
	with $r=1+\frac n2$. The first step is to carry out the Feynman parametrization. We use
	\be
	\frac{1}{a_1 \cdots a_r b}= r! \int_0^1 dx_1\int_0^{1-x_1}dx_2\cdots \int_0^{1-\sum_{i=1}^{r-1} x_i} dx_r
	\frac{1}{\,\,\,\Big[\sum_{k=1}^r(a_k-b)x_k + b \Big]^{r+1}},
	\ee
	with $a_k=[(p_k-k)^2-m_k^2]$ and $b=(k^2-m^2)$. Considering the denominator is given by $D^{r+1}$, it is possible to rearrange $D$ in order to write
	\be
	D=\left[k - \sum_{k=1}^r p_k x_k\right]^2 + Q^2,
	\ee
	with
	\be
	Q^2=\sum_{k=1}^r\left[p_k^2 x_k(1-x_k) + (m^2-m_k^2)x_k \right] 
	- \sum_{k \neq l}(p_k\cdot p_l)x_k x_l -m^2
	\ee
	We now perform a shift in the integral in the momentum: $k \to k + q$, with $q_\mu = \left[\sum_{k=1}^r p_k x_k\right]_\mu$, and we get, in the numerator,
	\be
	N_{\mu_1 \cdots \mu_n}= \prod_{i=1}^n (k+q)_{\mu_i},
	\ee
	from which only the even powers of $k$ survive, let us call it $\tilde N_{\mu_1 \cdots \mu_n}$. Our logarithmic divergent amplitude is then given by
	\be
	I^{(0)}_{\mu_1 \cdots \mu_n}=r! \int dX \int_k^R \frac{\tilde N_{\mu_1 \cdots \mu_n}}{\,\,\,[k^2+Q^2]^{r+1}},
	\label{int-geral}
	\ee
	where $\int dX$ stands for all the integrals in the Feynman parameters. The higher power in $k$ in $\tilde N_{\mu_1 \cdots \mu_n}$ is responsible for the logarithmic divergence. All the other terms are finite. For the divergent part, we have
	\bq
	I^{(0)\Lambda}_{\mu_1 \cdots \mu_n} &=& r! \int dX \, \int_k^R \frac{k_{\mu_1} \cdots k_{\mu_n}}{\,\,\,(k^2+Q^2)^{r+1}} \nonumber \\
	&=& \frac{\eta_{\mu_1 \cdots \mu_n}}{2^{\frac n2} }  \int dX\,  
	\int_k^R\frac{1}{(k^2+Q^2)^2},
	\eq
	in which we have used the consistency relation of (\ref{RC-geral}) and the fact that $r=\frac n2 + 1$. Next, we use the the scale relation (\ref{scale-1}) to obtain
	\bq
	I^{(0)\Lambda}_{\mu_1 \cdots \mu_n}= \frac{\eta_{\mu_1 \cdots \mu_n}}{2^{\frac n2}} \int dX\,
	\left\{ I_{log}(m^2)- \frac{i}{16 \pi^2}\ln{\left(-\frac{Q^2}{\lambda^2}\right)} \right\}.
	\eq
	The unique part which is dependent of the Feynman parameters is the one that contains $Q^2$. The other can be factorized out of the integral. The final result is given by
	\be
	I^\Lambda_{\mu_1 \cdots \mu_n}= \frac{\eta_{\mu_1 \cdots \mu_n}}{2^{\frac n2}} 
	\left\{ \frac{1}{r!}I_{log}(m^2) - \frac{i}{16 \pi^2} Z^{(r,0)}\right\},
	\ee
	in which we define the function
	\be
	Z_{\mu_1 \cdots \mu_i}^{(r,k_1,\cdots,k_r)} = Z_{\mu_1 \cdots \mu_i}^{(r,k_1,\cdots,k_r)}(p_1, \cdots, p_r, m_1^2, \cdots , m_r^2, m^2)
	\equiv \int dX\, q_{\mu_1} \cdots q_{\mu_i}x_1^{k_1}\cdots x_r^{k_r} \ln{\left(-\frac{Q^2}{m^2}\right)}.
	\ee
	We now have to take care of the other finite terms, which appears if $n \geq 2$. A typical finite term of (\ref{int-geral}) is
	\be
	F^{(l)}_{\mu_1 \cdots \mu_n} = r! \int dX\, {\cal S}\left[A_{\mu_1 \cdots \mu_l} q_{\mu_{l+1}} \cdots q_{\mu_n}\right],
	\ee
	with $l < n$ even and
	\be
	A_{\mu_1 \cdots \mu_l} = \int_k \frac{k_{\mu_1} \cdots k_{\mu_l}}{\,\,\,(k^2+Q^2)^{r+1}} =
	\frac{i}{16 \pi^2} \frac{\Gamma(\frac{n-l}{2})}{2^{\frac l2}\Gamma(r+1)}
	\frac{1}{\,\,\,(Q^2)^{\frac{n-l}{2}}}  \eta_{\mu_1 \cdots \mu_l}.
	\ee
	We, then, stay with
	\bq
	F^{(l)}_{\mu_1 \cdots \mu_n} &=& \frac{i}{16 \pi^2} \frac{\Gamma(\frac{n-l}{2})}{2^{\frac l2}}\,  {\cal S}\left[\eta_{\mu_1 \cdots \mu_l}\, \int dX\, \frac{q_{\mu_{l+1}} \cdots q_{\mu_n}}{(Q^2)^{\frac{n-l}{2}}}\right] \nonumber \\
	&=& \frac{i}{16 \pi^2}\frac{\Gamma(\frac{n-l}{2})}{2^{\frac l2}} {\cal S}\left[\eta_{\mu_1 \cdots \mu_l} Y_{\mu_{l+1} \cdots \mu_n}^{\left(r,\frac{n-l}{2}\right)} \right],
	\eq
	where
	\be
	Y_{\mu_1 \cdots \mu_n}^{(r,l)}= Y_{\mu_1 \cdots \mu_n}^{(r,l)}(p_1, \cdots, p_r,m_1^2, \cdots, m_r^2, m^2)
	\equiv  \int dx_1 \cdots dx_r\, \frac{q_{\mu_1} \cdots q_{\mu_n}}{(Q^2)^l}.
	\ee
	Finally, we can write the general result as
	\bq
	I^{(0)}_{\mu_1 \cdots \mu_n} &=& \frac{\eta_{\mu_1 \cdots \mu_n}}{2^{\frac n2}} 
	\left\{ \frac{1}{r!}I_{log}(m^2)- \frac{i}{16 \pi^2} Z^{(r,0)}\right\} +
	\nonumber \\
	&+& \frac{i}{16 \pi^2} \sum_{i=0}^{\frac{n-2}{2}} \frac{\Gamma(\frac{n-2i}{2})}{2^i} {\cal S}\left[\eta_{\mu_1 \cdots \mu_{2i}} Y_{\mu_{2i+1} \cdots \mu_n}^{\left(r,\frac{n-2i}{2}\right)} \right]; \,\,\,\,\,\,\,\,\,\,\,\, r = \frac n2 + 1,
	\eq
	in which we substituted $l$ by $2i$, since $l$ is even.
	
	For a linearly divergent integral,
	\be
	I^{(1)}_{\mu_1 \cdots \mu_n}=\int_k^R \frac{k_{\mu_1}\cdots k_{\mu_n}}{(k^2-m^2)[(p_1-k)^2-m_1^2]
		\cdots [(p_r-k)^2-m_r^2]},
	\ee
	with  $n$ odd and $r=\frac {n+1}{2}$, the procedure is very similar, with the result
	\bq
	I^{(1)}_{\mu_1 \cdots \mu_n} &=& \frac{1}{2^{\frac{n-1}{2}}}{\cal S}\left[ \sum_{k=1}^r p_{k\mu_1}\eta_{\mu_2 \cdots \mu_n} \right]
	\frac{1}{(r + 1)!}I_{log}(m^2)
	- \frac{i}{16 \pi^2} \frac{1}{2^{\frac{n-1}{2}}} {\cal S}\left[Z_{\mu_1}^{(r,0)}\eta_{\mu_2 \cdots \mu_n}\right] +
	\nonumber \\
	&+& \frac{i}{16 \pi^2} \sum_{i=0}^{\frac{n-3}{2}} \frac{\Gamma(\frac{n-2i-1}{2})}{2^{(2i+1)/2}} {\cal S}\left[\eta_{\mu_1 \cdots \mu_{2i}} Y_{\mu_{2i+1} \cdots \mu_n}^{\left(r,\frac{n-2i-1}{2}\right)} \right];
	\,\,\,\,\,\,\,\,\,\,\,\, r = \frac{n+1}{2},
	\eq
	in which the last term only appears for $n \geq 3$.
	
	In order to complete the systematization of the calculation of one-loop amplitudes, we turn our attention now to quadratically divergent integrals, $I^{(2)}_{\mu_1 \cdots \mu_n}$. Since the calculation is a little more involved, we will show below the main steps. The integral to be calculated has the same form as (\ref{I-zero}), but with $r=\frac n2$, $n$ even. After Feynman parametrization, we obtain the expression of (\ref{int-geral}), which will be broken in three parts: the higher power in $k$ in the numerator is quadratically divergent; the $(n-2)$-th power in $k$ in the numerator is logarithmically divergent; and the $(n-4)$-th power and lower, if they exist, are finite. It is important to note that even the two divergent peaces contribute to the finite result, as it is evident from the calculations of the last section. 
	
	Let us begin with the quadratic divergence:
	\bq
	I^{(2),1}_{\mu_1 \cdots \mu_n}=\left(\frac n2 \right)! \int dX \, \int_k^R \frac{k_{\mu_1} \cdots k_{\mu_n}}{\,\,\,(k^2+Q^2)^{\frac n2 + 1}}
	= \frac{\eta_{\mu_1 \cdots \mu_n}}{2^{\frac n2} }  \int dX\, 
	\int_k^R\frac{1}{(k^2+Q^2)},
	\eq
	in which we made use of (\ref{RC-geral-2}). Next, we resort to the scale relation of equation (\ref{scale-2}) to obtain
	\bq
	&& I^{(2),1}_{\mu_1 \cdots \mu_n}=\frac{\eta_{\mu_1 \cdots \mu_n}}{2^{\frac n2} }  \int dX\, \left\{ I_{quad}(m^2) - (m^2+Q^2)\left[I_{log}(m^2) 
	+ \frac{i}{16 \pi^2}\right] + \frac{i}{16 \pi^2} Q^2 \ln{\left(-\frac{Q^2}{m^2} \right)} \right\} \nonumber \\
	&& = \frac{\eta_{\mu_1 \cdots \mu_n}}{2^{\frac n2} } \left\{ \frac{1}{\left(\frac n2 \right)!}I_{quad}(m^2) 
	- \frac{1}{\left(\frac n2 +2 \right)!}\left\{ \sum_{k=1}^r \left[\frac n2 p_k^2 + \left(\frac n2 +2\right)(m^2 -m_k^2) \right] + \right. \right. \nonumber \\
	&& \left. \left. - \sum_{k \neq l}(p_k \cdot p_l)\right\}\left[I_{log}(m^2) + \frac{i}{16 \pi^2}\right]
	+ \frac{1}{16 \pi^2} \int dX \, Q^2 \ln{\left(-\frac{Q^2}{m^2} \right)} \right\},
	\eq
	where the last integral can be written in terms of $Z^{(r,k_1,\cdots,k_r)}$ functions.
	
	For the second part, which is logarithmically divergent, we have
	\bq
	&& I^{(2),2}_{\mu_1 \cdots \mu_n}=\left(\frac n2\right)!  \int dX\, {\cal S}\left[ q_{\mu_1}q_{\mu_2}
	\int_k \frac{k_{\mu_3} \cdots k_{\mu_n}}{\,\,\,(k^2+Q^2)^{\frac n2 +1}} \right] \nonumber \\
	&& = \frac{1}{2^{\frac{n-2}{2}}} \int dX\, {\cal S} [q_{\mu_1}q_{\mu_2} \eta_{\mu_3 \cdots \mu_n}]
	\left\{ I_{log}(m^2)- \frac{i}{16 \pi^2} \ln{\left(-\frac{Q^2}{m^2} \right)} \right\} \nonumber \\
	&& = \frac{1}{2^{\frac{n-2}{2}}}{\cal S}\left[\eta_{\mu_3 \cdots \mu_n} 
	\left\{ \frac{1}{\left(\frac n2 + 2\right)!} \left[ 
	2\sum_{k=1}^{n/2}p_{k\mu_1}p_{k\mu_2} + \sum_{k \neq l}p_{k\mu_1}p_{l\mu_2}\right]
	I_{log}(m^2) - \frac{i}{16 \pi^2} Z^{(n/2,0)}_{\mu_1 \mu_2} \right\}\right].
	\eq
	
	The remaining finite part is given by
	\bq
	&& I^{(2),3}_{\mu_1 \cdots \mu_n}=\left(\frac n2\right)! \sum_{l=4}^{n} \int dX\, {\cal S}\left[ q_{\mu_1} \cdots q_{\mu_l}
	\int_k \frac{k_{\mu_{l+1}} \cdots k_{\mu_n}}{\,\,\,(k^2+Q^2)^{\frac n2 +1}} \right] \nonumber \\
	&& = \frac{i}{16 \pi^2} \sum_{l=4}^{n} \frac{\Gamma\left(\frac l2 -1\right)}{2^{(n-l)/2}} \int dX\, {\cal S}\left[ q_{\mu_1} \cdots q_{\mu_l} \eta_{\mu_{l+1}\cdots \mu_n}\right]
	\frac{1}{(Q^2)^{\frac l2 -1}} \nonumber \\
	&& = \frac{i}{16 \pi^2} \sum_{i=2}^{n/2} \frac{\Gamma( i -1)}{2^{(n-2i)/2}}
	{\cal S}\left[\eta_{\mu_{2i+1} \cdots \mu_n} Y_{\mu_1 \cdots \mu_{2i}}^{(m,i-1)} \right].
	\eq
	The total result for the quadratically divergent integral is given by $I^{(2)}_{\mu_1 \cdots \mu_n}=I^{(2),1}_{\mu_1 \cdots \mu_n}+I^{(2),2}_{\mu_1 \cdots \mu_n}+I^{(2),3}_{\mu_1 \cdots \mu_n}$.
	
	\section{Directions on higher order calculations}
	
	There is currently a high demand for theoretical predictions for processes at next-to-next-to-leading order (NNLO) and beyond, mainly due to the large amount of data which has already been collected at LHC. With this aim, new calculation techniques have been developed in recent years, which seek, as far as possible, to preserve the physical dimension of the spacetime \cite{NNLO}. Higher-loop calculations are usually very long and intricate and friendly procedures are welcome. There are very nice techniques to implement multiloop calculations in quantum field theory, mainly with the use of Dimensional Regularization (see, for example, \cite{Grozin}). In the context of this new approach for implementing constrained IReg, we give some directions for future systematization of multi-loop calculations. Before starting, we note that it is not always possible, in higher-order calculations, to make a single Feynman parametrization for the entire amplitude, as in the case of non-planar graphs. However, as long as shifts in the momenta of integration are allowed in all steps, momentum routing invariance will be ensured.
	
	In the last section, we obtained general finite parts which are integrals, in the Feynman parameters, which contain factors of $\ln{[Q^2/(-m^2)]}$ or powers of $1/Q^2$. The challenging part is the one which includes the logarithm. With the aim of applying Feynman parametrization, we will make use of the following mathematical identity:
	\be
	\ln{a} = \lim_{\varepsilon \to 0} \frac {1}{\varepsilon} (a^\varepsilon - 1). 
	\ee
	In a simple view, we say that $\ln{a}$ is equal to the first order coefficient of the expansion of $a^\varepsilon$ in the limit of small $\varepsilon$. This trick in the context of Feynman integrals at higher loop orders, as far as we know, was first used in \cite{ir11} and \cite{irn} to calculate finite parts.
	
\begin{figure}[h]
		\begin{center}
			\includegraphics[scale=0.6]{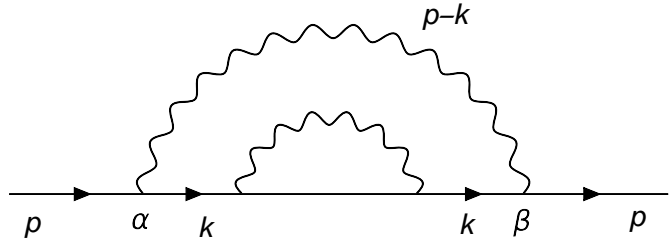}
			\caption{Diagrammatic representation of the nested contribution to the electron self-energy at two-loop order. The wavy and solid lines represent the photon and fermion propagators, respectively.}
			\label{Fig3}
		\end{center}
	\end{figure}

	So, let us give an example in which the one-loop finite part is an integral of a function with a factor of a logarithm. We consider the two-loop nested self-energy of the electron in QED, which is depicted in Figure \ref{Fig3}. 
	
	The finite part of the subgraph is given by
	\be
	i \tilde{\Sigma}^{(1)}= 2 q^2 \frac{i}{16 \pi^2}\int_0^1 dx \, [2m - \pslash (1-x)]\ln{\left(-\frac{H^2}{m^2}\right)},
	\ee
	with $H^2=H^2(p^2,m^2)=p^2x(1-x)-m^2x$. The integral for the two-loop graph is then written as
	\be
	i \Sigma^{(2)}(p)=-2iq^4 \frac{i}{16 \pi^2}\int_0^1 dx \,\int_k^R \frac{\gamma^\alpha(\kslash + m)[2m + (x-1)\kslash](\kslash + m)\gamma_\alpha}{(k^2-m^2)^2(p-k)^2}\ln{\left(-\frac{H^2(k^2,m^2)}{m^2}\right)}.
	\ee
	Let us define
	\be
	F=-2iq^4 \frac{i}{16 \pi^2}\int_0^1 dx \,\int_k^R \frac{\gamma^\alpha(\kslash + m)[2m + (x-1)\kslash](\kslash + m)\gamma_\alpha}{(k^2-m^2)^2(p-k)^2}\left(-\frac{H^2}{m^2}\right)^\varepsilon,
	\ee
	such that $i \Sigma^{(2)}(p)=F_\varepsilon$, being $F_\varepsilon$ the first order coefficient in the expansion of $F$ in powers of $\varepsilon$. We then write
	\be
	\left(-\frac{H^2}{m^2}\right)^\varepsilon = \left[\frac{x(1-x)}{(-m^2)}\right]^\varepsilon \, (k^2 - \tilde{m}^2)^\varepsilon; \,\,\,\, \tilde{m}^2=\frac{m^2}{(1-x)},
	\ee
	to obtain
	\be
	F=-2iq^4 \frac{i}{16 \pi^2}\int_0^1 dx \,\left[\frac{x(1-x)}{(-m^2)}\right]^\varepsilon\int_k^R \frac{\gamma^\alpha(\kslash + m)[2m + (x-1)\kslash](\kslash + m)\gamma_\alpha(k^2 - \tilde{m}^2)}{(k^2-m^2)^2(k^2 - \tilde{m}^2)^{1-\varepsilon}(p-k)^2}.
	\ee
	In the equation above, we multiplied the numerator and the denominator by a factor of $(k^2 - \tilde{m}^2)$ for convenience to obtain a well defined Feynman parametrization, from which, we obtain
	\be
	F=-2iq^4 \frac{i}{16 \pi^2}\frac{\Gamma(4-\varepsilon)]}{\Gamma(1-\varepsilon)}\int_0^1 dx \,\int_0^1 du \, \int_0^{1-u}dv \, v \left[\frac{x(1-x)}{(-m^2)(1-u-v)}\right]^\varepsilon\int_k^R \frac{\tilde{N}}{(k^2+Q^2)^{4-\varepsilon}},
	\ee
	where $Q^2=p^2u(1-u)- \tilde{m}^2(1-u-v)-m^2v$ and $\tilde{N}$ is the even part in $k$ of
	\be
	N=\gamma^\alpha(\kslash + \pslash u + m)[2m + (x-1)(\kslash + \pslash u)](\kslash + \pslash u + m)\gamma_\alpha[(k + pu)^2 - \tilde{m}^2].
	\ee
	
	The divergent part of $F$ is the one with quartic terms in $k$ in the numerator. Let us carry out explicitly the calculation of this part. After performing the Dirac algebra, the quartic part of the numerator is written as
	\bq
	N^{(4)}&=& 2k^4[4mx - u(x-1)\pslash] - 8 k^2 u(x-1) (p \cdot k) \kslash \nonumber \\
	&=& 2(k^2+Q^2)^2 [4mx - u(x-1)\pslash] -4(k^2+Q^2) Q^2 [4mx - u(x-1)\pslash]  \nonumber \\
	&+& 2Q^4 [4mx - u(x-1)\pslash] - 8(k^2+Q^2) u(x-1) (p \cdot k) \kslash + 8Q^2 u(x-1) (p \cdot k) \kslash,
	\eq
	in which we added and subtracted $Q^2$ to the factors of $k^2$. The first and fourth terms will result in logarithmically divergent integrals. For the first, we have
	\bq
	F^{(4)}_1= -4iq^4 \frac{i}{16 \pi^2}\frac{\Gamma(4-\varepsilon)]}{\Gamma(1-\varepsilon)}\int_{x,u,v} v \left[\frac{x(1-x)}{(-m^2)(1-u-v)}\right]^\varepsilon  [4mx - u(x-1)\pslash]
	\int_k^R \frac{1}{(k^2+Q^2)^{2-\varepsilon}},
	\eq
	with $\int_{x,u,v}$ representing the integrals in the Feynman parameters. We then expand the above expression for small $\varepsilon$ to get the coefficient of the first order term
	\bq
	F^{(4)}_{1 \varepsilon}&=&-4iq^4 \frac{i}{16 \pi^2}\int_{x,u,v} v [4mx - u(x-1)\pslash] \left\{ 
	6I_{log}^{(2)}(-Q^2,m^2) + \right. \nonumber \\
	&+& \left. \left[6 \ln{\left[\frac{x(1-x)}{(1-u-v)}\right]}-11 \right] I_{log}(-Q^2) \right\},
	\eq
	in which, we use of the definition of the basic logarithmic divergence typical of two-loop order,
	\be
	I_{log}^{(2)}(m^2,\lambda^2)= \int_k^R \frac{1}{(k^2-m^2)^2}\ln{\left[\frac{(k^2-m^2)}{(-\lambda^2)}\right]}.
	\ee
	Besides the scale relations (\ref{scale-1}) and (\ref{scale-2}) we have already used for the one-loop calculations, we can easily obtain, through the same algebraic manipulations, the corresponding two-loop one,
	\be
	I^{(2)}_{log}(m^2, \lambda^2)= I^{(2)}_{log}(\lambda^2) - \frac{i}{16 \pi^2}\left\{\ln{\left(\frac{m^2}{\lambda^2}\right)} + \frac 12 \ln^2{\left(\frac{m^2}{\lambda^2}\right)} \right\},
	\label{scale-2loop}
	\ee
	where $I^{(2)}_{log}(\lambda^2) \equiv I^{(2)}_{log}(\lambda^2,\lambda^2)$. This relation will be used to obtain a divergent part free from the external momenta. But let us first get the result of the other divergent term,
	\be
	F^{(4)}_4=16 i q^4 \frac{i}{16 \pi^2} \frac{\Gamma(4-\varepsilon)]}{\Gamma(1-\varepsilon)} \int_{x,u,v} vu(x-1) \left[\frac{x(1-x)}{(-m^2)(1-u-v)}\right]^\varepsilon 
	p^\alpha \gamma^\beta \int_k^R \frac{k_\alpha k_\beta}{\,\,(k^2+Q^2)^{3-\varepsilon}}.
	\ee
	For the integral in $k$, we can write
	\be
	\int_k^\Lambda \frac{k_\alpha k_\beta}{(k^2+Q^2)^{3-\varepsilon}}=\frac{1}{2(2-\varepsilon)}\left\{\int_k^R   \frac{\eta_{\alpha \beta}}{\,\,(k^2+Q^2)^{2-\varepsilon}} - \int_k^R  \frac{\partial}{\partial k^\beta} \frac{k_\alpha}{\,\,(k^2+Q^2)^{2-\varepsilon}}\right\},
	\ee
	from which we discard the surface term as prescribed by constrained IReg. After substituting the above relation in $F^{(4)}_4$ and picking the first order coefficient of the expansion in powers of $\varepsilon$, we get
	\be
	F^{(4)}_{4\varepsilon}=8 i q^4 \frac{i}{16 \pi^2} \pslash \int_{x,u,v} vu(x-1) \left\{ 
	3I_{log}^{(2)}(-Q^2,m^2) + \left[3 \ln{\left[\frac{x(1-x)}{(1-u-v)}\right]}-4 \right] I_{log}(-Q^2) \right\}.
	\ee
	We now get together the two divergent integrals, use the scale relations  (\ref{scale-1}) and (\ref{scale-2loop}) and integrate the coefficients of the basic divergences to obtain
	\bq
	F^{(4)}_{1\varepsilon} + F^{(4)}_{4\varepsilon} &=& -\frac i2 q^4 \frac{i}{16 \pi^2} \left\{2(\pslash + 8m)I_{log}^{(2)}(m^2) - (3 \pslash + 32m)I_{log}(m^2)\right\} + \nonumber \\
	&+& 4 i q^4 \left( \frac{i}{16 \pi^2}\right)^2 \int_{x,u,v}v \left\{\left\{ 4mx  \left[6 \ln{\left[\frac{x(1-x)}{(1-u-v)}\right]}- 5 \right] + \right. \right. \nonumber \\
	&-& \left. \left.  u(x-1) \pslash  \left[12 \ln{\left[\frac{x(1-x)}{(1-u-v)}\right]} - 7 \right]\right\} \ln{\left(-\frac{Q^2}{m^2}\right)} + \right.
	\nonumber \\
	&+& \left. 6 [2mx - u(x-1) \pslash] \ln^2{\left(-\frac{Q^2}{m^2}\right)}.
	\right\}
	\eq
	
	There are still three finite integrals considering the quartic part of the numerator. Besides, we have finite integrals coming from the quadratic and zeroth order in $k$ terms in the numerator. These calculations are straightforward: the integration in $k$ is performed, resulting in $\varepsilon$-depending powers of $Q^2$ in the denominator; the expansion in powers of $\varepsilon$ is performed to take the first order coefficient. The final result for the two-loop amplitude is given by
	\bq
	i\Sigma^{(2)}(p)&=& -\frac i2 q^4 \frac{i}{16 \pi^2} \left\{2(\pslash + 8m)I_{log}^{(2)}(m^2) - (3 \pslash + 32m)I_{log}(m^2)
	+ \frac{1}{72}\frac{i}{16 \pi^2} (5 \pslash + 352 m)  \right\} + \nonumber \\
	&+&  4 i q^4 \left( \frac{i}{16 \pi^2}\right)^2 \int_{x,u,v} v \left\{ A \ln{\left(-\frac{Q^2}{m^2}\right)} 
	\ln{\left[-\frac{Q^2}{m^2}\left(\frac{x(1-x)}{(1-u-v)}\right)^2\right]} + \right. \nonumber \\
	&+& \left.  \frac{1}{Q^2}  \left[B \ln{\left[-\frac{Q^2}{m^2}\frac{x(1-x)}{(1-u-v)}\right]} + C\right] + \frac{D}{Q^4}\left[  \ln{\left[-\frac{Q^2}{m^2}\frac{x(1-x)}{(1-u-v)}\right]} - 1\right] \right\},
	\eq
	with
	\bq
	A &=& 6 [2mx-u(x-1) \pslash], \\
	B &=& 2\left\{u \pslash [5u^2p^2(x-1)+2m^2(x+4))] -2m[2u(u+3)xp^2+3(m^2-x\tilde{m}^2)]\right\}, \\
	C &=& 4m(2uxp^2+m^2-\tilde{m}^2x) - u \pslash [2u^2p^2(x-1)+m^2(x+4)], \\
	D &=& -(u^2p^2-\tilde{m}^2)\left\{4m(xu^2p^2+m^2) -u \pslash [u^2(x-1)p^2 + m^2 (x+3)] \right\}.
	\eq
	
	There are some interesting comments about the two-loop calculation above. First, the basic divergence $I_{log}^{(2)}(m^2)$ appears naturally, even for massive models, when the expansion in $\varepsilon$ is carried out. Second, the momentum integration of the finite part is easily performed with the help of Feynman parametrization. And, last but not least, the elimination of the surface terms is simpler than the traditional procedure adopted in IReg \cite{irn}, which needs new relations for each loop-order. 
	\subsection{Massless models}
	The calculation above is just an example of how the present approach can be applied to a two-loop massive model, which is a calculation difficult to be implemented in the usual procedure of IReg. It is interesting to have some higher order results to be compared with the ones obtained before. Let us then resort to non-massive spinorial electrodynamics. For the two-loop graph treated above, the procedure is not simply the limit $m \to 0$, since the divergence of the one-loop subgraph was subtracted with the physical mass as the parameter. However, it is simple to carry out this calculation.
	
	For the one-loop subgraph, we have
	\be
	i\Sigma^{(1)}=-q^2 \int_k^R \frac{\gamma^\alpha \kslash \gamma_\alpha}{k^2(p-k)^2}
	=q^2 \pslash\left\{I_{log}(\lambda^2)- \frac{i}{16 \pi^2}\left[ \ln{\left(-\frac{p^2}{\lambda^2}\right)} - 2\right]\right\}
	\ee
	Subtracting out the one-loop subdivergence, we have for the two-loop nested diagram
	\be
	i \Sigma^{(2}(p)= - 2i q^4 \frac{i}{16 \pi^2} \int_k^R \frac{\kslash}{k^2(p-k)^2}
	\left[ \ln{\left(-\frac{k^2}{\lambda^2}\right)} - 2\right] = A_1 + A_2.
	\ee
	For $A_1$, the part with contains the logarithm, we use $A_1=F_\varepsilon$, as before, with
	\bq
	F&=& - 2i q^4 \frac{i}{16 \pi^2} (-\lambda^2)^{-\varepsilon} \int_k^R \frac{\kslash}{(k^2)^{1-\varepsilon}(p-k)^2} \nonumber \\
	&=& - 2i q^4 \frac{i}{16 \pi^2} (-\lambda^2)^{-\varepsilon} \frac{\Gamma(2-\varepsilon)}{\Gamma(1-\varepsilon)} \int_0^1 dx \, (1-x)^{-\varepsilon}
	\int_k^R \frac{\kslash +\pslash x}{\,\,\,(k^2+H^2)^{2-\varepsilon}}
	\eq
	
	After expanding for $\varepsilon \to 0$, collecting the first order coefficient and using the scale relations as prescribed, one obtains
	\bq
	A_1= iq^4\frac{i}{16 \pi^2} \pslash \left\{-I_{log}^{(2)}(\lambda^2) 
	- \frac 12 I_{log}(\lambda^2) + \frac 12 \frac{i}{16 \pi^2} 
	\left[\ln^2{\left(-\frac{p^2}{\lambda^2}\right)} - \ln{\left(-\frac{p^2}{\lambda^2}\right)} -3 \right]\right\}
	\eq
	
	The term $A_2$ is a simple one-loop integral. When the terms are put together, we get
	\be
	i \Sigma^{(2}(p)= iq^4\frac{i}{16 \pi^2} \pslash \left\{-I_{log}^{(2)}(\lambda^2) 
	+ \frac 32 I_{log}(\lambda^2) + \frac 12 \frac{i}{16 \pi^2} 
	\left[\ln^2{\left(-\frac{p^2}{\lambda^2}\right)} - 5 \ln{\left(-\frac{p^2}{\lambda^2}\right)} + 5 \right]\right\},
	\ee
	which can be compared with the result of \cite{irn}, where higher order calculations with IReg in massless gauge theories have been implemented. In that paper, the complete renormalization of spinorial QED to two-loop order has been performed. 
	
	For a more significant comparison, let us look at the photon self-energy. The diagrams which contribute are exhibited in Figure 4.
	\begin{center}
		\begin{picture}(300,100)(0,20)
		\Photon(20,50)(50,50){4}{3} \Photon(110,50)(140,50){4}{3}
		\Photon(54.02,65)(105.98,65){3}{5}
		\CArc(80,50)(30,0,360)  \Text(80,10)[t]{$(a)\,\,\,\,-i\Pi^{\mu\nu}_a(p)$} 
		\Photon(160,50)(190,50){4}{3} \Photon(250,50)(280,50){4}{3} \Photon(220,20)(220,80){3}{5}
		\CArc(220,50)(30,0,360)  \Text(220,10)[t]{$(b)\,\,\,\,-i\Pi^{\mu\nu}_b(p)$}
		\end{picture} 
		\vspace{1cm}
		\\ FIG. 4. Contributions to the two-loop photon self-energy in spinorial QED.
	\end{center}
	
	The amplitude for the graph of Figure 4$(a)$, corresponding to the nested contribution, can be computed by simply using the finite part of the subgraph, which is the electron self-energy. We have
	\begin{eqnarray}
	-i \Pi^{\mu\nu}_a = -2i\frac{i}{16 \pi^2}q^4\int_k^R 
	\frac{\mbox{tr}\left\{\gamma^\nu \kslash \gamma^\mu (\kslash - \pslash) \right\}}{k^2 (k-p) ^2} \left[ \ln{\left(-\frac{p^2}{\lambda^2}\right)} - 2\right],
	\label{nested-TPV}
	\end{eqnarray}
	where the factor $2$ is to take into account the graph with the insertion in the inferior line of the loop. The second term is proportional to the one-loop vacuum polarization tensor. For the first one, which contains a logarithm, we write
	\begin{equation}
	-i\Pi^{\mu\nu}_{a1F} = -2i \frac{i}{16 \pi^2}\frac{q^4}{(-\lambda^2)^\varepsilon}\int_k^R 
	\frac{\mbox{tr}\left\{\gamma^\nu \kslash \gamma^\mu (\kslash - \pslash) \right\}}{(k^2)^{1-\varepsilon} (k-p) ^2},
	\end{equation}
	from which we will extract the coefficient of the first power in $\epsilon$. We Feynman parametrize the complete integral to obtain
	\begin{equation}
	-i \Pi^{\mu\nu}_{a1F} = -8i\frac{i}{16 \pi^2}q^4\frac{\Gamma(2-\varepsilon)}{\Gamma(1-\varepsilon)} \int_0^1 dx [-(1-x)\lambda]^{-\varepsilon}\int_k^R \frac{\left[2k^\mu k^\nu - k^2 \eta^{\mu\nu} + x(1-x) (p^2 \eta^{\mu\nu} - 2 p^\mu p^\nu)\right]}{[k^2 + H^2]^{2-\varepsilon}},
	\end{equation}
	with $H^2=p^2x(1-x)$. It is possible to reorganize the terms so as to have, in the numerator of the $k$ integral,
	\begin{equation}
	2 k^\mu k^\nu - (k^2 + H^2) \eta^{\mu\nu} + 2x(1-x) (p^2 \eta^{\mu\nu} -p^\mu p^\nu).
	\end{equation}
	The two first terms will originate quadratically divergent integrals. Since
	\begin{equation}
	\frac{\partial}{\partial k_\nu} \left( \frac{k^\mu}{(k^2+H^2)^{1-\varepsilon}}\right) = 
	\frac{\eta^{\mu\nu}}{(k^2+H^2)^{1-\varepsilon}} - 2(1-\varepsilon) \frac{k^\mu k^\nu}{(k^2+H^2)^{2-\varepsilon}},
	\end{equation}
	we can reduce the tensorial integral to a scalar one by discarding the surface term, so that we have, for these two terms,
	\begin{equation}
	- 8i\frac{i}{16 \pi^2}q^4\frac{\varepsilon\Gamma(2-\varepsilon)}{(1-\varepsilon)\Gamma(1-\varepsilon)} \int_0^1 dx [-(1-x)\lambda]^{-\varepsilon}\int_k^R \frac{\eta^{\mu\nu}}{\;\;\;[k^2 + H^2]^{1-\varepsilon}}.
	\end{equation}
	The coefficient of the first power in the expansion in $\varepsilon$ is then given by
	\begin{equation}
	-8i\frac{i}{16 \pi^2}q^4 \eta^{\mu\nu} \int_0^1 dx \; I_{quad}(-H^2).
	\end{equation}
	For the other part, we stay with
	\begin{equation}
	-16i\frac{i}{16 \pi^2}q^4 (p^2 \eta^{\mu\nu} -p^\mu p^\nu) \frac{\Gamma(2-\varepsilon)}{\Gamma(1-\varepsilon)} \int_0^1 dx [-(1-x)\lambda]^{-\varepsilon} x(1-x)\int_k^R \frac{1}{\;\;\;[k^2 + H^2]^{2-\varepsilon}},
	\end{equation}
	which furnishes, for the first order coefficient,
	\begin{equation}
	-16i\frac{i}{16 \pi^2}q^4 (p^2 \eta^{\mu\nu} -p^\mu p^\nu) \int_0^1 dx \; x(1-x)
	\left[I_{log}^{(2)}(-H^2) - [\ln (1-x) + 1] I_{log}(-H^2) \right].
	\end{equation}
	Using the scale relations for one- and two-loop, we obtain
	\begin{eqnarray}
	&& -16i\frac{i}{16 \pi^2}q^4 (p^2 \eta^{\mu\nu} -p^\mu p^\nu) \int_0^1 dx \; x(1-x)
	\left\{I_{log}^{(2)}(\lambda^2) -\frac 12 \frac{i}{16 \pi^2} \ln^2\left(-\frac{H^2}{\lambda^2}\right) - \frac{i}{16 \pi^2} \ln\left(-\frac{H^2}{\lambda^2}\right) + \right. \nonumber \\
	&& \left. -[\ln(1-x) + 1] \left[ I_{log}(\lambda^2) - \frac{i}{16 \pi^2} \ln\left(-\frac{H^2}{\lambda^2}\right)\right]\right\} \nonumber \\
	&&= -\frac 83 i \frac{i}{16 \pi^2}q^4(p^2 \eta^{\mu\nu} -p^\mu p^\nu) \left\{I^{(2)}_{log}(\lambda^2) - \frac 16 I_{log}(\lambda^2) - \frac{i}{16 \pi^2} \left[ \frac 12  
	\ln^2\left(-\frac{p^2}{\lambda^2}\right) + \frac 56 
	\ln\left(-\frac{p^2}{\lambda^2}\right) \right]\right\}.
	\end{eqnarray}
	We still have the second term of eq. (\ref{nested-TPV}), which gives
	\begin{equation}
	-i \Pi^{\mu\nu}_{a2} = -\frac 83 i \frac{i}{16 \pi^2}q^4(p^2 \eta^{\mu\nu} -p^\mu p^\nu) \left\{ -2 I_{log}(\lambda^2) + \frac{i}{16 \pi^2} \left[\ln\left(-\frac{p^2}{\lambda^2}\right) - \frac{10}{3}\right]\right\},
	\end{equation}
	to be added to the first one to give, for the nested graph,
	\begin{eqnarray}
	-i\Pi^{\mu\nu}_{a} &=& -\frac 83 i\frac{i}{16 \pi^2}q^4(p^2 \eta^{\mu\nu} -p^\mu p^\nu) \left\{ I^{(2)}_{log}(\lambda^2) - \frac{13}{6} I_{log}(\lambda^2) + \frac{i}{16 \pi^2} \left[- \frac 12  
	\ln^2\left(-\frac{p^2}{\lambda^2}\right) + \right. \right. \nonumber \\
	&+& \left. \left. \frac 16\ln\left(-\frac{p^2}{\lambda^2}\right) - \frac{10}{3}\right]\right\} 
	- 8 i \frac{i}{16 \pi^2}q^4 \eta^{\mu\nu} \int_0^1 dx \; I_{quad}(-H^2).
	\end{eqnarray}
	
	Note that the unique term which violates the transversality of the amplitude is the one in $I_{quad}(-H^2)$. This part, if the appropriated scale relations is used, encompasses several contributions in addition to the quadratic divergence. We opt to maintain them gathered in this way, since they must be cancelled out when added to the result for the overlapped diagram.
	
	The amplitude for the overlapped graph of Fig. 4$(b)$, with the addition of the counterterms to cancel out the subdivergences, is given by
	\bq
	-i \Pi^{\mu\nu}_{b} &=&  2iq^4 \int_k^R \int_l^R \frac{\mbox{tr}\{\gamma^\nu \lslash(\kslash-\pslash)\gamma^\mu \kslash (\lslash - \pslash)\}}{k^2l^2(k-l)^2(p-k)^2(p-l)^2} + \nonumber \\
	&+& \frac 83 iq^2 I_{log}(\lambda^2)\left[ I_{log}(\lambda^2) - \frac{i}{16 \pi^2} \ln\left(-\frac{p^2}{\lambda^2}\right) + \frac 53 \frac{i}{16\pi^2}\right]
	(p^2 \eta^{\mu\nu} - p^\mu p^\nu).
	\eq
	It is important to note that, usually, the counterterms to cancel the subdivergences are calculated as different graphs to be added to the amplitude, as prescribed by the forest formula of BPHZ. Here, equivalently, we simply defined a quantity which is already free from the subdivergences, just to give a treatment similar to the one given to the nested diagram, for which we used only the finite part of the subgraph. 
	
	For the overlapped graph, it is simpler to break the amplitude in a combination of integrals and then apply, to these integrals, the proposed approach. After the calculation of the trace of the Dirac matrices, we obtain, for the numerator (using the symmetry in exchanging $k$ and $l$),
	\bq
	N^{\mu\nu} &=& 8iq^4\left\{ k^2 l^2 \eta^{\mu\nu} -4 k^2 l^\mu l^\nu + 4 (k \cdot l) k^\mu k^\nu + 4k^2 l^\nu p^\mu + 2p^2l^\mu k^\nu - 2k^2 (l \cdot p) \eta^{\mu\nu} + \right.\nonumber \\
	&-& \left. 4 (k \cdot l) l^\nu p^\mu - 4 (k \cdot p) l^\mu p^\nu - 4 (k \cdot p) k^\mu l^\nu + 4 (k \cdot p) l^\mu l^\nu + 2 (k \cdot p)(l \cdot p) \eta^{\mu\nu} + \right. \nonumber \\ 
	&-& \left. p^2 (k \cdot l) \eta^{\mu\nu} + 2 (k \cdot l) p^\mu p^\nu 
	\right\}.
	\label{numerator}
	\eq
	We will focus, in the above amplitude, in discussing some sensible points, like the cancellation of the quadratic divergences. The first three terms in eq. (\ref{numerator}) originate quadratically divergent integrals, which should give us the exact term in order to cancel out the gauge violating remaining term in the nested diagram. Let us calculate the first one. We have
	\be
	J^{\mu\nu}=8iq^4 \eta^{\mu\nu} \int_k^R \int_l^R \frac{k^2 l^2}{k^2 l^2 (k-l)^2 (p-k)^2 (p-l)^2},
	\ee
	in which we cancel only the factor $k^2$, that is sufficient to make easy the solution of the integral. It is important to note that, by performing appropriated shifts, the dependence of $J^{\mu\nu}$ on the external momentum $p$ is completely eliminated. However, we intend to maintain this dependence in different parts of the result (obviously, this dependence cancels out when the parts are put together) The integration in $k$ gives
	\be
	I_{log}(\lambda^2) + \frac{i}{16 \pi^2} \left[ 2 - \ln\left[-\frac{(p-l)^2}{\lambda^2}\right] \right].
	\ee
	We then stay with
	\be
	J^{\mu\nu} = 8iq^4 \eta^{\mu\nu} \left\{ \left[I_{log}(\lambda^2) + 2 \frac{i}{16 \pi^2}\right] \int_l^R \frac{l^2}{l^2(p-l)^2} - \frac{i}{16 \pi^2}  \int_l^R \frac{l^2}{l^2(p-l)^2}
	\ln\left[-\frac{(p-l)^2}{\lambda^2}\right] \right\}.
	\ee
	For the second integral, we use the procedure to reduce the logarithm to powers of $\epsilon$. For both integrals, we apply Feynman parametrization and, only after this step, we cancel the factors of $l^2$ by adding and subtractig $H^2$. In the end, we obtain
	\bq
	J^{\mu\nu} &=& \left[I_{log}(\lambda^2) + 2 \frac{i}{16 \pi^2}\right] \int_0^1 dx \, I_{quad}(-H^2) 
	+ \nonumber \\
	&-& \frac{i}{16\pi^2} \int_0^1 dx \left\{ I_{quad}^{(2)}(-H^2,\lambda^2)- [\ln(1-x)+1] I_{quad}(-H^2)\right\} + \cdots,
	\eq
	with
	\be
	I_{quad}^{(2)}(m^2,\lambda^2)= \int_k^R \frac{1}{(k^2-m^2)}\ln{\left[\frac{(k^2-m^2)}{(-\lambda^2)}\right]}
	\ee
	and in which we only displayed the terms that involve quadratic divergences. The second quadratically divergent integral reads
	\bq
	A^{\mu\nu} &=& - 32 i q^4 \int_k^R \int_l^R \frac{k^2 l^\mu l ^\nu}{k^2 l^2 (k-l)^2 (p-k)^2 (p-l)^2} \nonumber \\
	&=& - 16 i q^4 \left\{  \left[I_{log}(\lambda^2) + 2 \frac{i}{16 \pi^2}\right] \int_0^1 dx \, I_{quad}(-H^2) \right.
	+ \nonumber \\
	&-& \left.\frac{i}{16\pi^2} \int_0^1 dx \left\{ I_{quad}^{(2)}(-H^2,\lambda^2)- \ln(1-x) I_{quad}(-H^2)\right\}\right\} + \cdots,
	\eq
	and, the third,
	\bq
	B^{\mu\nu} &=&  32 i q^4 \int_k^R \int_l^R \frac{k^\mu l ^\nu (k \cdot l)}{k^2 l^2 (k-l)^2 (p-k)^2 (p-l)^2} \nonumber \\
	&=& 8 i q^4 \left\{  \left[I_{log}(\lambda^2) + 2 \frac{i}{16 \pi^2}\right] \int_0^1 dx \, I_{quad}(-H^2) \right.
	+ \nonumber \\
	&-& \left.\frac{i}{16\pi^2} \int_0^1 dx \left\{ I_{quad}^{(2)}(-H^2,\lambda^2)- \ln(1-x) I_{quad}(-H^2)\right\} \right\} + \cdots.
	\eq
	Again, we only show the terms that contribute to quadratic divergences. The sum of these three terms gives
	\be
	J^{\mu\nu} + A^{\mu\nu} + B^{\mu\nu}= 8 i \frac{i}{16 \pi^2} q^4 \eta^{\mu\nu} 
	\int_0^1 dx \, I_{quad}(-H^2) + \cdots,
	\ee
	which is exactly the necessary to cancel out the quadratic divergence coming from the nested diagram. We should comment on a notable feature of this new procedure with regard to quadratic divergences. Usually in IReg, when dealing with massles models, quadratic divergences are disregarded on the grounds that it is possible to construct a parametrization in which they are null. Here it is not even necessary to use the scale relations for such divergences, as the cancellation occurs for the packages gathered in $I_{quad}(-H^2)$ and $I_{quad}^{(2)}(-H^2,\lambda^2)$.
	
	The other integrals originated from the terms of (\ref{numerator}) are easily calculated with the approach presented in this paper. The final result for the sum $-i\Pi^{\mu\nu}_a -i \Pi^{\mu\nu}_b$ agrees perfectly with the one of \cite{irn} and is given by
	\begin{eqnarray}
	\label{eq: pimunutot}
	-i\Pi^{\mu\nu} &=& \frac{8}{3}i q^4 \frac{i}{16 \pi 2}(p^\mu p^\nu -
	\eta^{\mu\nu}p^2)\left\{\frac{3}{2}I^2_{log}(\lambda^2) + \right. \nonumber \\
	&+& \left. \frac{i}{16\pi^2}\left[ - 3I^{(2)}_{log}(\lambda^2) + \frac{31}{6}I_{log}(\lambda^2) - \frac{3}{2}\ln\left(-\frac{p^2}{\lambda^2}\right)
	+ \frac{3}{2}\right] - p^2 I^{O}\right\},
	\end{eqnarray}
	with
	\be
	I^{O} = \int_k \int_l \frac{1}{k^2 l^2 (k-l)^2 (p-k)^2 (p-l)^2}.
	\ee
	As a final comment in this section, it is important to note that the use of Implicit Regularization beyond one-loop order is not new. For example, in \cite{ir8} the procedure was used to calculate the beta-function of the Wess-Zumino model in order of three loops. In the paper \cite{multiloop}, procedures were developed for the application of the technique in non-massive models in n-loop order, with the determination of scale and consistency relations for an arbitrary loop order.
	
	We believe that the procedure we discuss in this section improves the methodology of calculation and can be systematized and be of great help in multiloop phenomenological calculations.
	
	\section{Comment on Implicit Regularization of Infrared Divergent Amplitudes}
	
	Although the present study is focused in the regularization of ultraviolet divergences, we carry out below a brief discussion on the application of Implicit Regularization in the treatment of infrared divergences. This was the subject of the paper \cite{IRed}, in which one- and higher-loop orders applications were discussed. The procedure was used in the analysis of the origin of the two-loop contributions to $N=1$ super Yang-Mills beta-function \cite{2-loop}.
	
	The essential steps of Implicit Regularization, when used in the case of ultraviolet divergences, have their counterpart for infrared ones. Moreover, a new scale appears, typically an infrared scale which is completely independent of the ultraviolet one. In order to perform all the algebraic manipulations which are necessary in IReg, we assume the presence of a infrared regulator, which will be indicated by the upper index $\tilde{R}$. Considering that the integral is under this regularization, all the steps indicated in section \ref{new-approach} are allowed, such as using algebraic identities to separate the divergent part, performing shifts in the integration coordinate, with the consequent elimination of surface terms, and the cancellation of common factors in the numerator and denominator.

	First, we consider the following ultraviolet divergent massless integral and its result within Implicit Regularization,
	\be
	\label{ultradiv}
	I = \int^R\frac{d^{4}k}{(2\pi)^{4}}\frac{1}{k^{2}(p-k)^{2}}      = I_{log}(\lambda^{2})- \frac{i}{16\pi^2}\left[\ln\left(-\frac{p^{2}}{\lambda^{2}}\right) - 2\right],
	\ee
	and then proceed with the calculation of
	\be\label{infra}
	U = \int^{\tilde{R}}\frac{d^{4}k}{(2\pi)^{4}}\frac{1}{k^{4}(p-k)^{2}},
	\ee
	which, by power counting, is infrared divergent and ultraviolet finite. In order to be able to use all the tools developed for ultraviolet divergent integrals we firstly note that
	\be\label{kquatro}
	\frac{1}{k^{4}}=-\int d^{4}u\ e^{iku}\int^{\tilde{R}}\frac{d^{4}z}{(2\pi)^{4}}\frac{1}{z^{2}(z-u)^{2}},
	\ee
	in which $z$ and $u$ are configuration variables.
	
	Note the striking similarity between the above $z$ integral and equation (\ref{ultradiv}). We can thus write the result immediately:
	\begin{eqnarray}
	I(u^{2})=\int^{\tilde{R}} \frac{d^4z}{(2\pi)^4}\frac{1}{z^{2}(z-u)^{2}}
	= I_{log}(\tilde{\lambda}^{-2})-\frac{i}{16\pi^2}\left[\ln\left(-u^{2}\tilde{\lambda}^{2}\right)-2\right].\label{infrabdi}
	\end{eqnarray}
	But now $I_{log}(\tilde{\lambda}^{-2})$ is an infrared basic divergent integral and a scale relation
	has been used to introduce the infrared scale $l^2=1/ \tilde{\lambda}^2$.
	
	Using this result and (\ref{kquatro}) in (\ref{infra}) we have
	\begin{eqnarray}
	&&U = -\int_{k}^{\tilde{R}}\frac{1}{(p-k)^{2}}\int d^{4}u\ e^{iku}I(u^{2})
	=-\frac{i}{(4\pi)^{2}}\int_{k}^{\tilde{R}}\int d^{4}u\int d^{4}x\frac{e^{i(p-k)x}}{x^{2}}e^{iku}I(u^{2})\nonumber\\
	&&=-\frac{i}{(2\pi)^{2}}\int d^{4}u\ \frac{e^{ipu}}{u^{2}}\left\{\tilde{I}_{log}(\tilde{\lambda}^{-2})-\frac{i}{16\pi^2}\left[\ln\left(-u^{2}\tilde{\lambda}^{2}\right)-2\right]\right\}\nonumber\\
	&&=-\frac{1}{p^{2}}\left\{I_{log}(\tilde{\lambda}^{-2})+\frac{i}{16\pi^2}\left[\ln\left(-\frac{p^{2}}{\bar{\tilde{\lambda}}^{2}}\right)+2\right]\right\}\label{diviv},
	\end{eqnarray}
	with $\bar{\tilde{\lambda}}^2 \equiv \frac{4}{e^{2\gamma}}\tilde{\lambda}^{2}$, where $\gamma = 0,5772...$ is the Euler-Mascheroni constant.
	
	The technique also allows the treatment of integrals in which infrared and ultraviolet divergences are present at the same time. Let us consider, just as a toy example,
	\be
	G=\int^{R,\tilde{R}}\frac{d^{4}k}{(2\pi)^{4}}\frac{1}{k^{4}}.
	\ee
	The key point in order to separate these two kinds of divergence is to multiply the numerator and the denominator by a factor $(p-k)^2$. We stay with
	\be
	G= p^2 U 
	+ \int^{R}\frac{d^{4}k}{(2\pi)^{4}}\frac{1}{k^2(p-k)^2} 
	- 2 p^\mu \int^{\tilde{R}}\frac{d^{4}k}{(2\pi)^{4}}\frac{1}{k^{4}
		(p-k)^2},
	\ee
	in which the first, already calculated, term is only infrared divergent, the second one is only ultraviolet divergent and the last one is finite. The final result is given by
	\be
	G=I_{log}(\lambda^2) + I_{log}(\tilde{\lambda}^{-2})
	+ \frac{i}{16 \pi^2}\left[\ln{\left(\frac{\lambda^2}{\bar{\tilde{\lambda}}^{2}}\right)} + 2\right].
	\ee
	
	The examples discussed above are cases which involves off-shell infrared divergences. Implicit Regularization has not a systematized prescription for the case of on-shell infrared divergences. Usually, since this kind of singularity should be cancelled out in physical calculations, these cases are dealt by provisionally assigning a small mass to the particle, $p^2=\mu^2$.
	
	There are some difficult situations like the one of
	\be
	F=\int_k \frac{1}{k^2(k-p_1)^2(k-p_1-p_2)^2(k-p_1-p_2-p_3)^2},
	\ee
	with $p_1^2=p_2^2=p_3^2=(p_1+p_2+p_3)^2=0$. This integral is off-shell finite, but encompasses an unavoidable infrared divergence in the on-shell limit. We illustrate below the difficulty of treating this kind of integral using the approach of Implicit Regularization.
	
	First, let us Feynman parametrize the factor,
	\begin{equation}
	\frac{1}{k^2(k-p_1)^2}= \int_0^1 dx \frac{1}{[(k-p_1 x)^2 + H^2]^2},
	\end{equation}
	with $H^2=p_1^2 x (1-x)$. Since $p_1^2=0$, we have
	\begin{equation}
	\frac{1}{k^2(k-p_1)^2}= \int_0^1 dx \frac{1}{(k-p_1 x)^4},
	\end{equation}
	which, in $F$, gives us
	\begin{equation}
	F = \int_0^1 dx \int_k^{\tilde{R}} \frac{1}{(k-p_1x)^4(k-p_1-p_2)^2(k-p_1-p_2-p_3)^2}.
	\end{equation}
	We then perform the shift $k \to k + p_1x$ and use the inverse Fourier transform of $1/k^4$ to write
	\begin{equation}
	F = - \int_0^1 dx \int_k^{\tilde{R}} \int d^4 z e^{i kz} I(z^2) \frac{1}{(k-q)^2(k-q-p_3)^2}; \;\;\;\; 
	q = (1-x)p_1 + p_2. 
	\end{equation}
	The new shift $k \to k + q$ gives us
	\begin{equation}
	F = - \int_0^1 dx \int_k^{\tilde{R}} \int d^4 z e^{i (k+q)z} I(z^2) \frac{1}{k^2(k-p_3)^2}.
	\end{equation}
	As we have done before for $p_1$, since $p_3^2=0$, we write
	\begin{equation}
	\frac{1}{k^2(k-p_3)^2} = \int_0^1 dy \frac{1}{(k-p_3 y)^4}
	\end{equation}
	and, then, make the shift $k \to k + p_3 y$ in the $k$-integral to obtain
	\begin{eqnarray}
	F &=& - \int_0^1 dx \int_0^1 dy  \int d^4z \int_k^{\tilde{R}} \frac{e^{i(k+q+p_3y)z}}{k^4} I(z^2) \nonumber \\
	&=& \int_0^1 dx \int_0^1 dy \int d^4z \int d^4u \int_k^{\tilde{R}} e^{i(k+q+p_3y)z} e^{iku} I(z^2) I(u^2) \nonumber \\
	&=& \int_0^1 dx \int_0^1 dy \int d^4z \int d^4u \,\, \delta^{(4)}(u+z) e^{i(q+p_3y)z}  I(z^2) I(u^2) \nonumber \\ 
	&=& \int_0^1 dx \int_0^1 dy \int d^4z \,\,  e^{i(q+p_3y)z} [I(z^2)]^2.
	\end{eqnarray}
	The integrals in the Feynman parameters $x$ and $y$  are easily solved. We have
	\begin{eqnarray}
	&&\int_0^1 dx \int_0^1 dy \,\, e^{i(q+p_3y)z} = \int_0^1 dx \int_0^1 dy \,\, e^{i[(1-x)p_1 + p_2 + p_3y]z} \nonumber \\
	&& = e^{ip_2 z}\left\{\int_0^1 dx \,\, e^{i(1-x)p_1 z}\right\} \left\{\int_0^1 dy \,\, e^{iyp_3 z}\right\} \nonumber \\
	&& = - \frac{e^{ip_2z}}{(p_1z)(p_3z)} \left[e^{ip_1z}-1\right]\left[e^{ip_3z}-1\right].
	\end{eqnarray}
	
	Finally, we can write
	\begin{eqnarray}
	F=-\int d^4z \frac{e^{ip_2z}}{(p_1z)(p_3z)} [e^{ip_1z}-1][e^{ip_3z}-1] 
	\left\{ I_{log}(\tilde{\lambda}^{-2})-\frac{i}{16\pi^2}\left[\ln\left(-z^{2}\tilde{\lambda}^{2}\right)-2\right]  \right\}^2.
	\end{eqnarray}
	In the equation above, we have a set of Fourier transforms calculated in $p_2$, $p_1+p_2$, $p_1+p_2+p_3$ and $p_2+p_3$. It is important to note that the infrared divergent content is enclosed in the $I_{log}(\tilde{\lambda}^{-2}))$ which is free from the external momentum. Nevertheless, it is clear that this approach of using integrals in the configuration space is not the best one to deal with on-shell infrared divergences. There are good strategies to treat on-shell infrared divergences, including higher order loops, like, for example, the one described in \cite{Anastasiou}.

	\section{Concluding comments}
	In this paper, we have established a new procedure for the application of Implicit Regularization in its constrained version. The constrained version of Implicit Regularization, which fixes all the surface terms to zero, automatically delivers symmetric amplitudes as already demonstrated in a wide variety of articles. This is due to the fact that symmetries such as gauge-invariance are related, in the context of Feynman integrals, to momentum routing invariance in the loops \cite{A-Vieira}. This new approach uses this fact with the aim of Feynman parametrize the complete amplitude after a regularization is assumed to be implicitly acting in the divergent integral. As it is well known, the usefulness of Feynman parametrization lies in the possibility of making a shift in the momentum of integration. A shift in a divergent integral would have to be compensated with a surface term if the degree of divergence is at least linear. This is why the amplitude should be regularized before Feynman parametrization is carried out. In addition, the regularization prescription must be such that these surface terms are null. This is the case of Dimensional Regularization, which turns the amplitude finite in the extended dimension and thus forces the surface terms to vanish. In the case of the constrained version of Implicit Regularization, this is accomplished with the help of the consistency relations.
	
	The procedure which is presented in this paper enforces momentum routing invariance by shifting the momentum of integration after the Feynman parametrization of the complete amplitude. It also fixes other remaining surface terms to zero. The great simplification in the approach occurs in consequence of the way the divergent part is separated from the finite one. While in the traditional application of IReg the integrand is expanded before Feynman parametrization, which is carried out only in the finite part, here this separation takes place after this step, by using scale relations that are always the same. We then avoid to deal with finite integrals with high powers in the momenta in the numerator and the denominator. Another advantage is the unification of the procedure to be adopted in massive and non-massive models, since the scale relations are in charge of introducing the mass parameter for the basic divergences and for the renormalization group equations. The great simplification in one-loop calculation for this new approach extends for higher-loop orders, as demonstrated in section V. The procedure above can be systematized and be of great help in multiloop phenomenological calculations. This is part of a future work.

	\section{Acknowledgements}
	
	This work was partially supported by Conselho Nacional de Desenvolvimento Cient\'ifico e Tecnol\'ogico (CNPq).

\end{document}